\documentclass{ptephy_v1}
\preprintnumber{XXXX-XXXX} 
\usepackage{subfig}

\usepackage{ulem}

\newcommand{\higashino}[1]{\textcolor{black}{#1}}

\begin{document}

\newtheorem{theorem}{Theorem}
\newtheorem{condition}{Condition}


\title{Direction-sensitive dark matter search with 
three-dimensional vector-type tracking in NEWAGE}


\author[1]{Takuya Shimada}
\author[1]{Satoshi Higashino}
\author[2]{Tomonori Ikeda}
\author[3]{Kiseki Nakamura}
\author[1]{Ryota Yakabe}
\author[1]{Takashi Hashimoto}
\author[1]{Hirohisa Ishiura}
\author[1]{Takuma Nakamura} 
\author[1]{Miki Nakazawa} 
\author[1]{Ryo Kubota}
\author[1]{Ayaka Nakayama}
\author[8]{Hiroshi Ito}
\author[6]{Koichi Ichimura}
\author[4,5]{Ko Abe}
\author[7]{Kazuyoshi Kobayashi}
\author[2]{Toru Tanimori}
\author[2]{Hidetoshi Kubo}
\author[2]{Atsushi Takada}
\author[4,5]{Hiroyuki Sekiya}
\author[4,5]{Atsushi Takeda}
\author[1]{Kentaro Miuchi}

\affil[1]{Department of Physics, Graduate School of Science, Kobe University, Rokkodai-cho, Nada-ku, Kobe-shi, Hyogo, 657-8501, Japan 
\email{higashino@phys.sci.kobe-u.ac.jp}
}
\affil[2]{Division of Physics and Astronomy Graduate School of Science, Kyoto University, Kitashirakawaoiwake-cho, Sakyo-ku, Kyoto-shi, Kyoto, 606-8502, Japan}
\affil[3]{Department of Physics, Tohoku University, Aramakiazaaoba 6-3, Aoba-ku, Sendai-shi, Miyagi, 980-8578, Japan}
\affil[4]{Kamioka Observatory, Institute for Cosmic Ray Research, the University of Tokyo, Higashi-Mozumi, Kamioka-cho, Hida-shi, Gifu, 506-1205, Japan}
\affil[5]{Kavli Institute for the Physics and Mathematics of the Universe (WPI), the University of Tokyo, 5-1-5 Kashiwanoha, Kashiwa-shi, Chiba, 277-8582, Japan}
\affil[6]{Research Center for Neutrino Science, Tohoku University, Sendai 980-8578, Japan}
\affil[7]{Waseda Research Institute for Science and Engineering, Waseda University, 3-4-1 Okubo, Shinjuku, Tokyo 169-8555, Japan}
\affil[8]{Department of Physics, Tokyo University of Science, 2641 Yamazaki, Noda-shi, Chiba, 278-8510, Japan}

\begin{abstract}
NEWAGE is a direction-sensitive dark matter search experiment with a three-dimensional tracking detector based on a gaseous micro time projection chamber.
A direction-sensitive dark matter search was carried out at Kamioka Observatory with a total live time of 318.0~days resulting in an exposure of 3.18~kg$\cdot$days.
A new gamma-ray rejection and a head-tail determination analysis were implemented for this work.
No significant non-isotropic signal from the directional analysis  was found and a 90\% confidence level upper limit on spin-dependent WIMP-proton cross section of 25.7~pb for WIMP mass of 150~GeV/$c^2$ was derived.
This analysis marked the most stringent upper limit in the direction-sensitive dark matter searches.

\end{abstract}

\subjectindex{Dark matter, WIMP, $\mu$TPC, NEWAGE}

\maketitle

\section{Introduction}
\label{sec:introduction}
Existence 
of the dark matter in the universe is nowadays widely believed because %
the dark matter naturally explains 
observational results in various scales of the universe.
Weakly Interactive Massive Particles (WIMPs), which are promising candidates of the dark matter, have been searched for by a number of direct search experiments pursuing for the nuclear recoil by WIMPs\,\cite{ARBEY2021103865}.
However, no conclusive evidence of the direct detection of WIMPs was obtained yet.

There are two possible characteristic signatures for the direct detection of the dark matter.
One is the annual modulation in the energy spectrum caused by the Earth's motion around the Sun. 
The modulation amplitude is expected to be a few percent\,\cite{BAUM2019262}.
The other is the directional non-isotropy of the nuclei recoils.
Since the Solar System is orbiting in the Milkyway Galaxy, the incoming direction of the dark matter is biased to the direction of the Solar System's motion.
The directional distribution of the nuclear recoil also has an asymmetry and this asymmetric ratio can be as large as tenfold in some cases\,\cite{PhysRevD.37.1353}.
Thus, the observation of the non-isotropic signal for the nuclear recoil direction distribution is expected to be a strong evidence for the dark matter detection.

NEWAGE (NEw generation WIMP search with an Advanced Gaseous tracker Experiment) is a direction-sensitive direct WIMP search experiment 
using a low-pressure gaseous micro Time Projection Chamber ($\mu$-TPC) 
for the detection of three-dimensional (3D) tracks of recoil nuclei. 
NEWAGE started direction-sensitive direct WIMP searches in an underground laboratory in 2007 and has updated the results since then. 
In 2020, head-tail determinations of the nuclear tracks were implemented and a limit by a vector-like tracking analysis was obtained (NEWAGE2020 results\,\cite{yakabe_ptep}).
In 2021, the limit was updated
by installing a low alpha ray emission rate detector called LA$\mu$-PIC\,\cite{hashimoto_nim,ikeda_ptep}.
Here the limit was obtained without the vector-like analysis (NEWAGE2021 results) because of the limited  statistics.
In this paper, we report a result of a direction-sensitive dark matter search with a new gamma-ray rejection cut and a vector analysis for 3D-tracks (3D-vector analysis) for a data 2.4 times larger than NEWAGE2021 results in total.



\section{Detector}
A gaseous time projection chamber, NEWAGE-0.3b'', was used for this study.
The detector overview is described in subsection~\ref{sec:detector}.
Energy calibration using alpha rays are discussed in subsection~\ref{sec:calibration}.
Event selections already implemented in our previous analysis are summarized in subsection~\ref{sec:standard_event_selection}.
An event selection newly-added for this work utilizing the track information for a better gamma-ray rejection is described in subsection~\ref{sec:new-selection}.
The reconstruction method of the 3D-vector tracks is explained in subsection~\ref{sec:head-tail} as the head-tail analysis.
Finally, the detector performances on the efficiencies and the angular resolution of the nuclear recoil are shown in subsections~\ref{sec:efficiency} and \ref{sec:angular-resolution}, respectively.

\subsection{NEWAGE-0.3b''}\label{sec:detector}
\begin{figure}[h]
    \centering
    \includegraphics[width=0.8\textwidth]{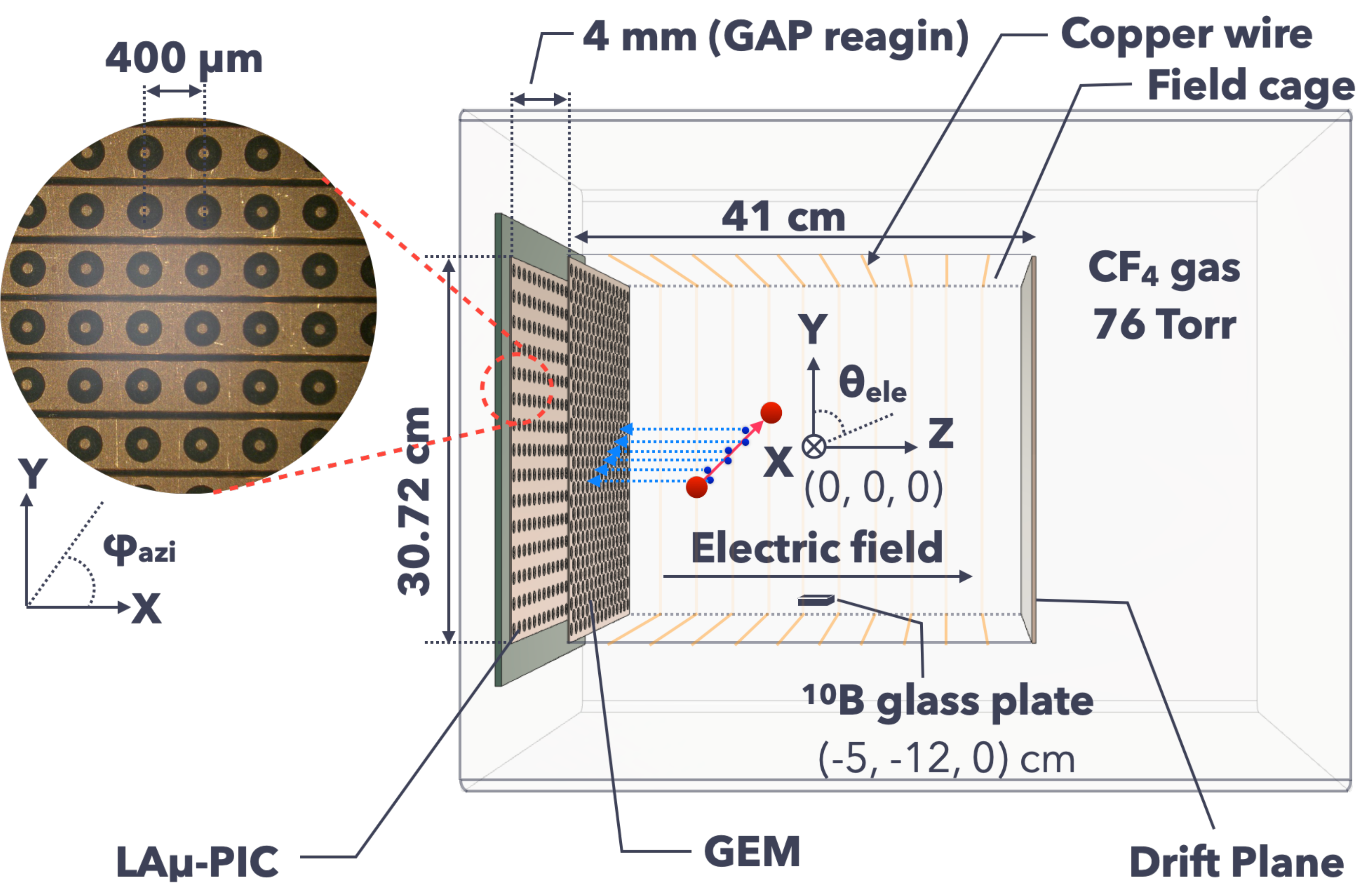}
    \caption{Schematic drawings of the NEWAGE-0.3b'' detector and its detection scheme.
    A recoil nucleus shown with red markers passes through the gas volume and ionizes the gas molecules (blue).
    The ionized electrons are drifted toward the readout plane by the electric field, amplified by the GEM\,\cite{GEM}, and further amplified by the LA$\mu$-PIC before being detected.
    The image on the left is a magnified view of the LA$\mu$-PIC with an  electrode structure of a 400~$\mu$m pitch. X, Y, and Z indicate the axes in the detector coordinate. $\phi_{\rm azi}$ and $\theta_{\rm ele}$ denote the azimuth and elevation angle of the detector coordinate, respectively.}
    \label{fig:detector}
\end{figure}

NEWAGE-0.3b'', refurbished in 2018 by replacing the 
readout device (micro pixel chamber, $\mu$-PIC) with a low alpha-emission rate one (LA$\mu$-PIC \,\cite{hashimoto_nim}), was used for this work.
Figure~\ref{fig:detector} shows schematic drawings of the NEWAGE-0.3b'' detector and its detection scheme. 
The detection volume was 31~$\times$~31$\times$~41~cm$^{3}$ in size and was filled with low-pressure gas of CF$_4$ at 76~Torr (0.1~atm) for this work.
The location of (0,~0,~0) in the detector coordinate is set at the center of the TPC.
The LA$\mu$-PIC has a pixel structure of 768~$\times$~768 with a pitch of 400~$\mu$m.
Amplified charge at each pixel is read through 768 anode (hereafter X-axis) and 768 cathode(hereafter Y-axis) strips.
Signals read through the strips are processed by  Amplifier-Shaper-Discriminator chips (SONY CXA3653Q\,\cite{ASD}).
The processed signals are then divided into two.
One is compared with a threshold voltage in the chips and the time-over-thresholds (TOTs) of 768 + 768 strips are recorded with a 100~MHz frequency clock.
The other 768 cathode strips were grouped into four channels and their waveforms were recorded with a 100~MHz flash analog-to-digital converters (FADCs).
A detected track is parameterized with its energy, length, elevation angle $\theta_{\rm ele}$, azimuth angle $\Phi_{\rm azi}$ (see Figure~\ref{fig:detector}) and some other parameters defined in the following subsections.

\subsection{Energy calibration}\label{sec:calibration}
\begin{figure}[h]
    \centering
    \includegraphics[width=0.6\textwidth]{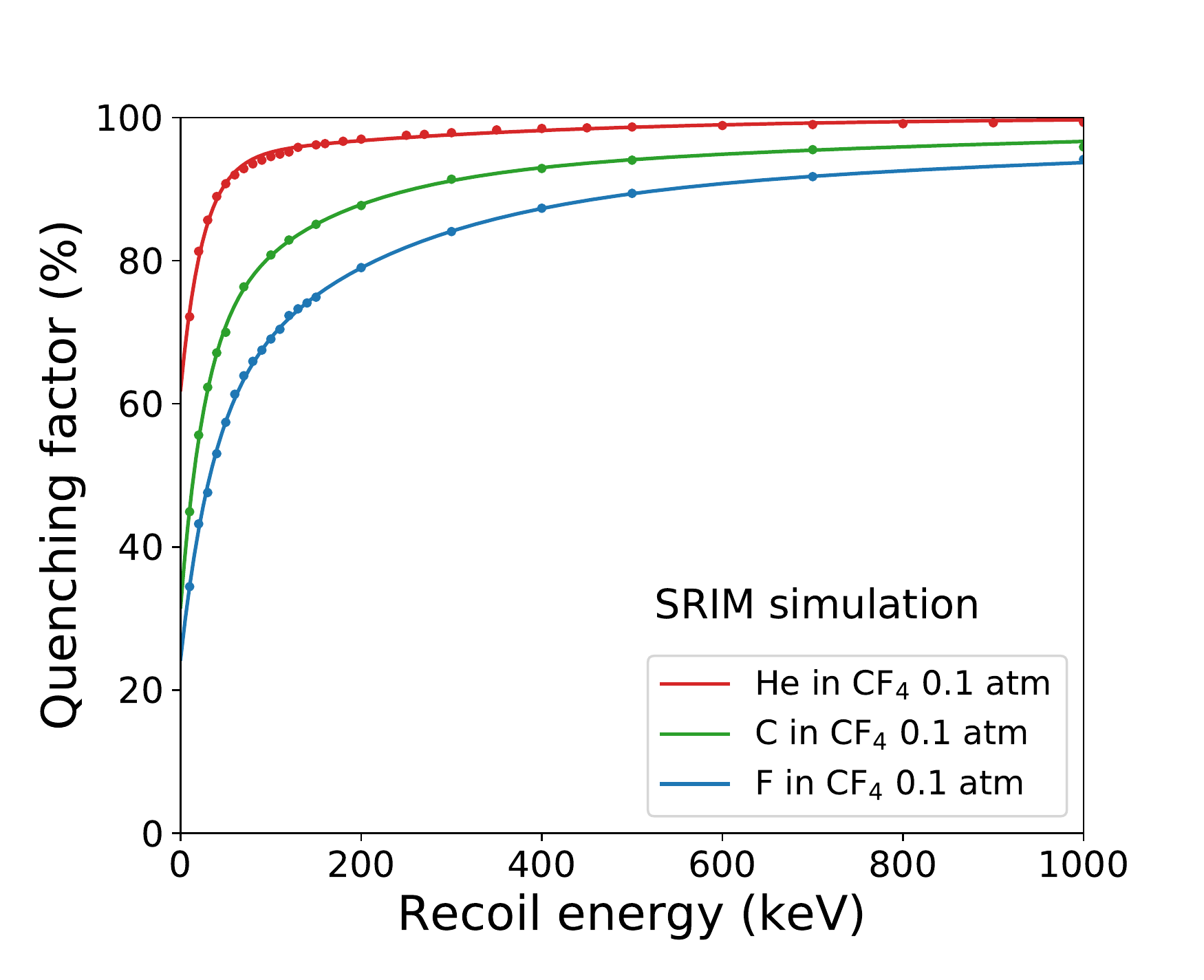}
    \caption{\higashino{Ionization quenching factors for He, C and F nuclei as a function of their recoil energy in the gas of 0.1~atm CF$_{4}$} calculated with SRIM~\cite{SRIM}.}
    \label{fig:qf}
\end{figure}

The energy calibration was performed with alpha rays produced by $\rm ^{10}B(n, \alpha){}^{7}Li$ reactions.
A glass plate coated with a $^{10}$B layer was set in the TPC volume
as illustrated in Fig.~\ref{fig:detector}.
Thermal neutrons were irradiated from the outside of the chamber, captured in the $^{10}$B layer, and then produced alpha rays and $^{7}$Li nuclei.
Because our $^{10}$B layer has a sub-micron thickness, alpha rays and $^{7}$Li nuclei deposit a part of energy in the $^{10}$B layer.
Consequently alpha rays and $^{7}$Li nuclei produce continuous spectrum up to 1.5~MeV and 0.8~MeV, respectively.
\higashino{Since the energy of the charged particle is not only converted to the ionization but partially deposited to other components such as phonon, it doesn't necessarily correspond to the detected energy. In order to take this effect, or the ionization quenching factor, into account, ionization quenching factors in CF$_{4}$ gas at 0.1 atm were calculated with SRIM~\cite{SRIM}. Figure~\ref{fig:qf} shows the ionization quenching factors for alpha-rays (He), C and F nuclei as functions of their recoil energies in CF$_{4}$ gas at 0.1 atm. In the following of this paper, the unit of detected energy is expressed as the electron equivalent energy ({\it i.e.} keV$_{ee}$), which takes the ionization quenching factor into account.}

\begin{figure}[h]
    \centering
    \includegraphics[width=0.6\textwidth]{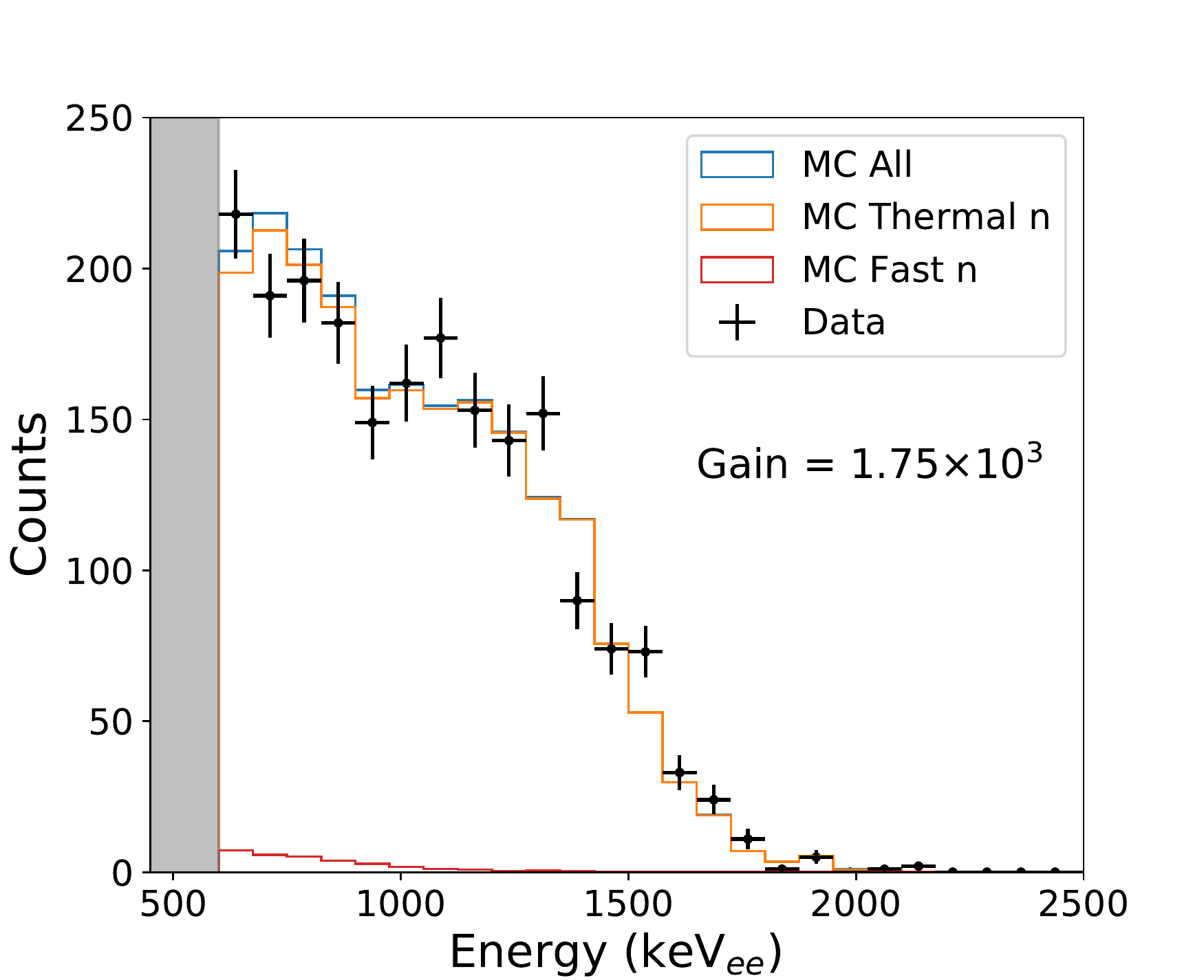}
    \caption{Energy spectrum of alpha rays from a $^{10}$B glass plate.
    The black plot is the measured data. The orange, red, and blue histograms are the simulated results for thermal neutrons, fast neutrons, and the sum of them, respectively.}
    \label{fig:calib}
\end{figure}

The obtained spectrum is a sum of the thermal neutron capture events and elastic scattering events by fast neutrons.
By comparing these spectra with the Monte-Carlo (MC) simulation results by Geant4\,\cite{AGOSTINELLI2003250}, the gas gain and the energy resolution were determined.
Figure~\ref{fig:calib} shows one of the calibration results. 
The 1.5~MeV and 0.8~MeV edges of the thermal neutrons were observed and consistent with our MC modelling.

\higashino{Although the energy calibration was performed using alpha-rays with an energy of $\sim$~1.5~MeV, the region of interest in energy is about 50~keV$_{ee}$.}
\higashino{In order to compensate this difference of energy scale, an additional study had been performed previously using smaller detector which had the same components as the NEWAGE-0.3''.}
\higashino{The study validate linearity of the energy scale using a $^{55}$Fe source in addition to the $^{10}$B plate, corresponding to the energy range between 5.9~keV and 1.5~MeV.}

The detector gas contains rare gas radon isotopes, $^{220}$Rn and $^{222}$Rn, emitted from the detector materials as natural contaminations.
The high-energy calibration was performed by the alpha rays from radon isotopes and  their progenies.
$^{220}$Rn and its progeny produce alpha rays with energies of 6.05~MeV, 6.29~MeV, 6.78~MeV, and 8.79~MeV, 
while $^{222}$Rn and its progeny produce alpha rays with energies of 5.49~MeV, 6.00~MeV, and 7.69~MeV.
Because the ratio of $^{220}$Rn to $^{222}$Rn was not known, the measured spectra were fit with 
the simulated spectrum of $^{220}$Rn and $^{222}$Rn separately
and the difference was treated as the systematic error of the energy scale.

\subsection{Standard event selections}
\label{sec:standard_event_selection}
\label{sec:conv-selection}
Several event selections had been established as standard event selections by NEWAGE2021 analysis.  
These selections aim to cut non-physical electronics noise events and electron track events mainly originating from ambient gamma-rays.
The standard event selections are briefly explained here, while  details can be found in Ref.\,\cite{ikeda_ptep}.

\begin{description}
\item[Fiducial volume cut]\mbox{}\\
A fiducial volume of 28~$\times$~24~$\times$~41~cm$^{3}$ was defined in 
the detection volume of 31~$\times$~31~$\times$~41~cm$^{3}$.
Any events were required to be fully contained in the fiducial volume so as to discriminate the events from the walls of the TPC field cage and the $^{10}$B glass plate.
%
\item[Length-Energy cut]\mbox{}\\
The amount of energy loss by a charged particle per a unit length depends on the particle type.
Electron events were discriminated by setting a maximum track length for a given energy.
%
\item[TOTsum/Energy cut]\mbox{}\\
The energy deposition on each strip 
was recorded as TOT.
A total TOTs of all strips were defined as TOTsum. 
Since the nuclear recoil events have larger TOTsum than those of the electron recoil events for a given energy, electron events were discriminated by setting a minimum 
TOTsum/energy value for a given energy.
(See the left panel of Fig.~\ref{fig:high-gain-bg}, for instance.)
%
\item[Roundness cut]\mbox{}\\
``Roundness'' was defined as the root-mean-square deviation of a track from the best-fit straight line. 
Nuclear recoil events with a short drift distance have small roundnesses because they are less affected by the gas diffusion.
Background events in the gas region between the LA$\mu$-PIC and the GEM were discriminated by setting a minimum roundness value.

\end{description}


\subsection{TOTsum-Length cut}
\label{sec:sec:additional_event_selection}
\label{sec:new-selection}
\begin{figure}[h]
    \centering
    \includegraphics[width=1.0\textwidth]{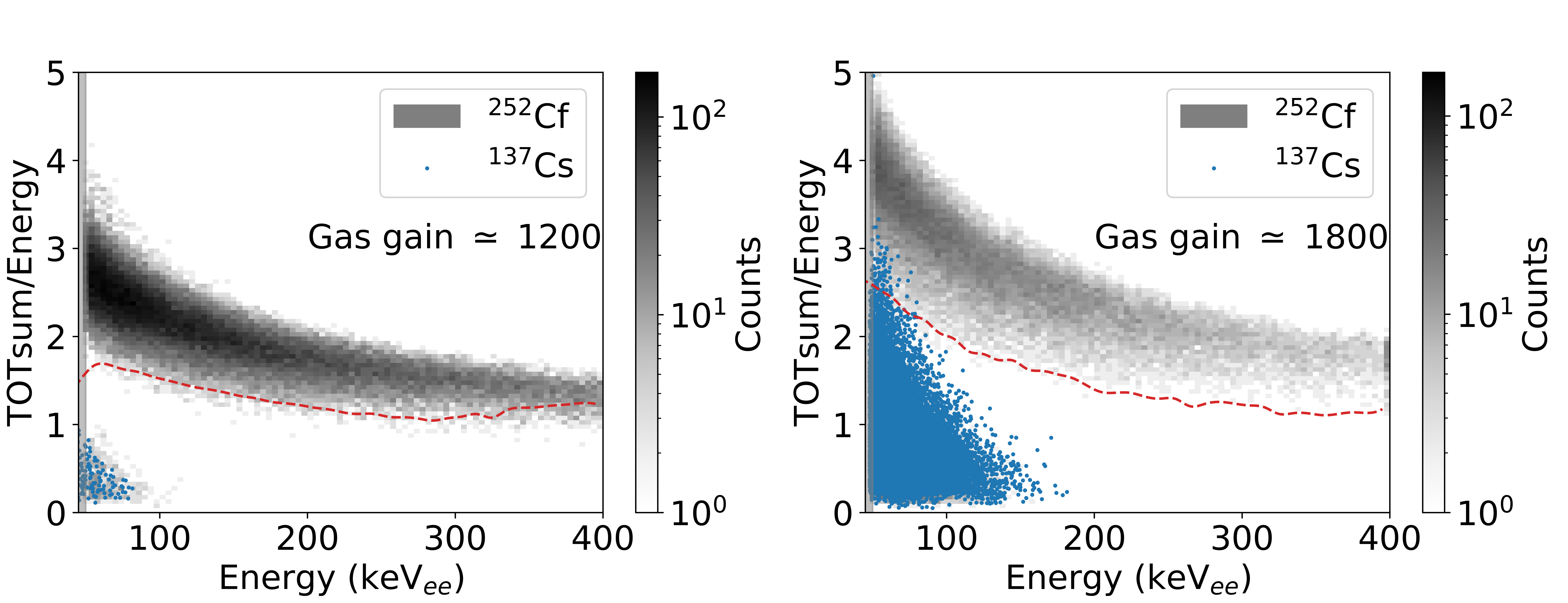}
    \caption{TOTsum/Energy distributions as functions of energy (after the fiducial volume cut). 
    The left and right panels show the distributions corresponding to the gas gain of 1200 and 1800, respectively.
    The black gradation distribution is obtained with a $^{252}$Cf neutron source.
    The blue point distribution is obtained with  a $^{137}$Cs gamma-ray source.
    The red-dashed lines indicate the cut lines.
    Each calibration run with the source had been conducted at a common live time of 0.18 days.}
    \label{fig:high-gain-bg}
\end{figure}

The detector was operated at a higher gas gain (typically 1800) than that of NEWAGE2021 (1200)
aiming for a better detection efficiency of nuclear recoil events. 
One of the expected drawbacks of the high-gain operation was the increase of 
the background gamma-ray events in contrast to the detection efficiency improvement of nuclear recoils owing to
the increase of the number of hit strips. 
Figure~\ref{fig:high-gain-bg} shows the TOTsum/Energy distributions as functions of the energy after the fiducial volume cut.
The gas gains of the left and right panels are 1200 and 1800, respectively.
It should be noted that each calibration run with the source had been conducted at a common live time of 0.18~days.
It is therefore clearly seen that the detection efficiency of electron events ($\rm ^{137}Cs$ data) are significantly larger in a measurement at a high gas gain because the number of shown events are increased. It is also seen that the TOTsum/Energy of electron events in the high-gain data have a large component which excess the selection line of TOTsum/Energy selection shown with a red line.
This result indicated that the standard event selections were not sufficient for the high-gain operation data. 

\begin{figure}[htbp]
	\begin{center}
		\begin{tabular}{c}
            \includegraphics[width=1.0\textwidth]{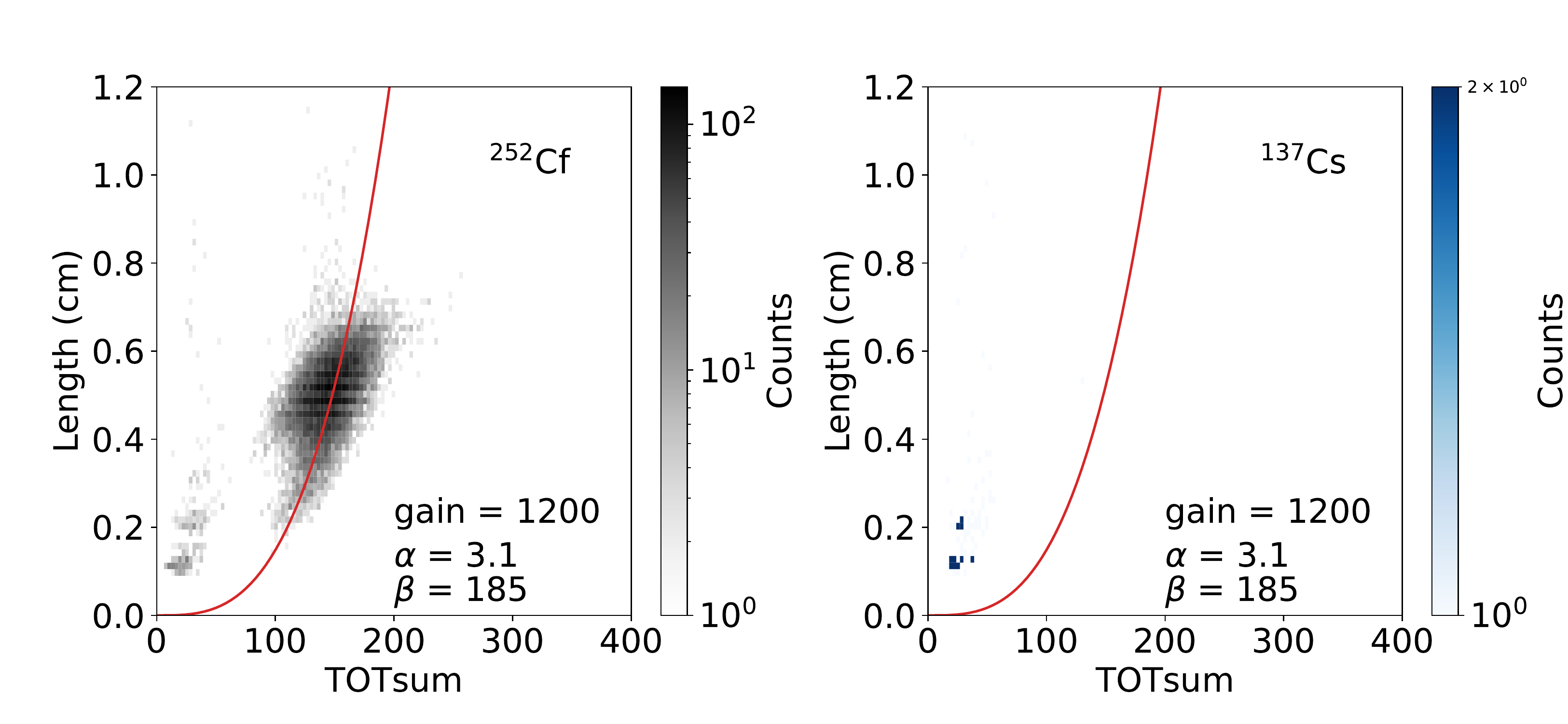}
			\hspace{1.6cm}
		\end{tabular}
		\begin{tabular}{c}
            \includegraphics[width=1.0\textwidth]{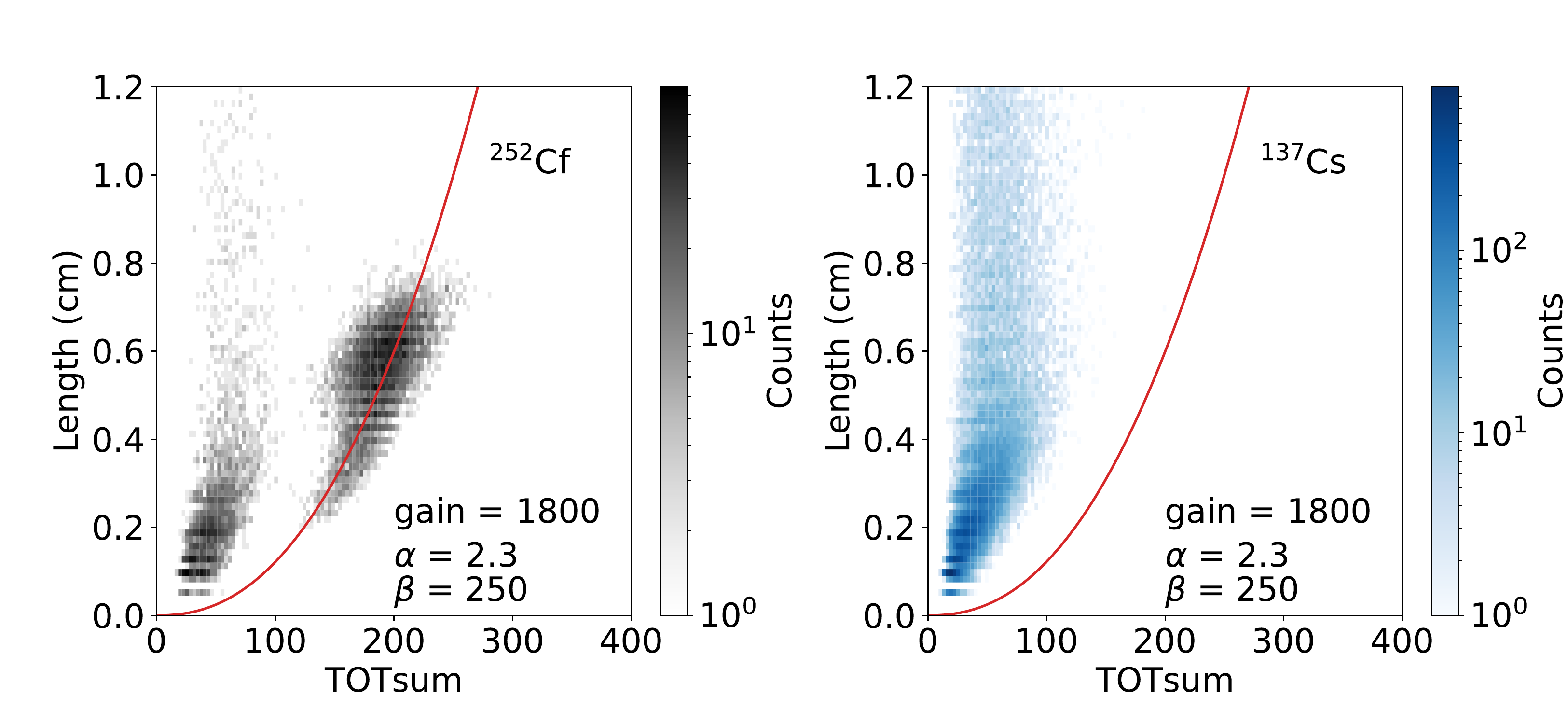}
			\hspace{1.6cm}
		\end{tabular}

        \caption{Distributions of track length as a function of TOTsum in the energy range of 50--60~keV$_{ee}$ after the fiducial volume cut. Upper and lower plots are the results measured with gas gains of 1200 and 1800, respectively.
        The left plots (black gradient) are the data with a $^{252}$Cf neutron source and the right (blue gradient) are the data with a $^{137}$Cs gamma-ray source.
        The red line in the figure is $\rm{L} = (\rm{S}/\beta)^\alpha$.
        Since the $^{252}$Cf source emits not only neutrons but also gamma-rays, the distribution has two components.}
        \label{fig:tot_length}
	\end{center}
\end{figure}

A new cut, ``TOTsum-Length cut'', was implemented in order to improve the discrimination power against the gamma-ray events. Nuclear recoil events have large TOTsums and short track lengths.
On the other hand, the electron recoil events have smaller TOTsums and longer tracks.
Figure~\ref{fig:tot_length} shows the track length distributions as a function of TOTsum for the irradiation with a $^{252}$Cf source and a $^{137}$Cs source for the cases with gas gains of 1200 and 1800.
Since our energy threshold is set to be 50~keV$_{ee}$, the data in an energy range of 50--60~keV$_{ee}$ are selected.
We confirmed a good separation of the electron (seen in both plots) and nuclear distributions (seen only in the $^{252}$Cf plot) in this parameter space even for a high-gain operation data. 
In order to discriminate electron events, an empirical function written by
\begin{equation}
    \rm{L} = (\rm{S}/\beta)^\alpha,
\end{equation}
was introduced.
Here L is the track length, S is the TOTsum, and $\alpha$ and $\beta$ are parameters for the cut definition. Here $\alpha$ was fixed within a run while $\beta$ was an energy-dependent parameter.

\begin{figure}[h]
    \centering
    \includegraphics[width=0.8\textwidth]{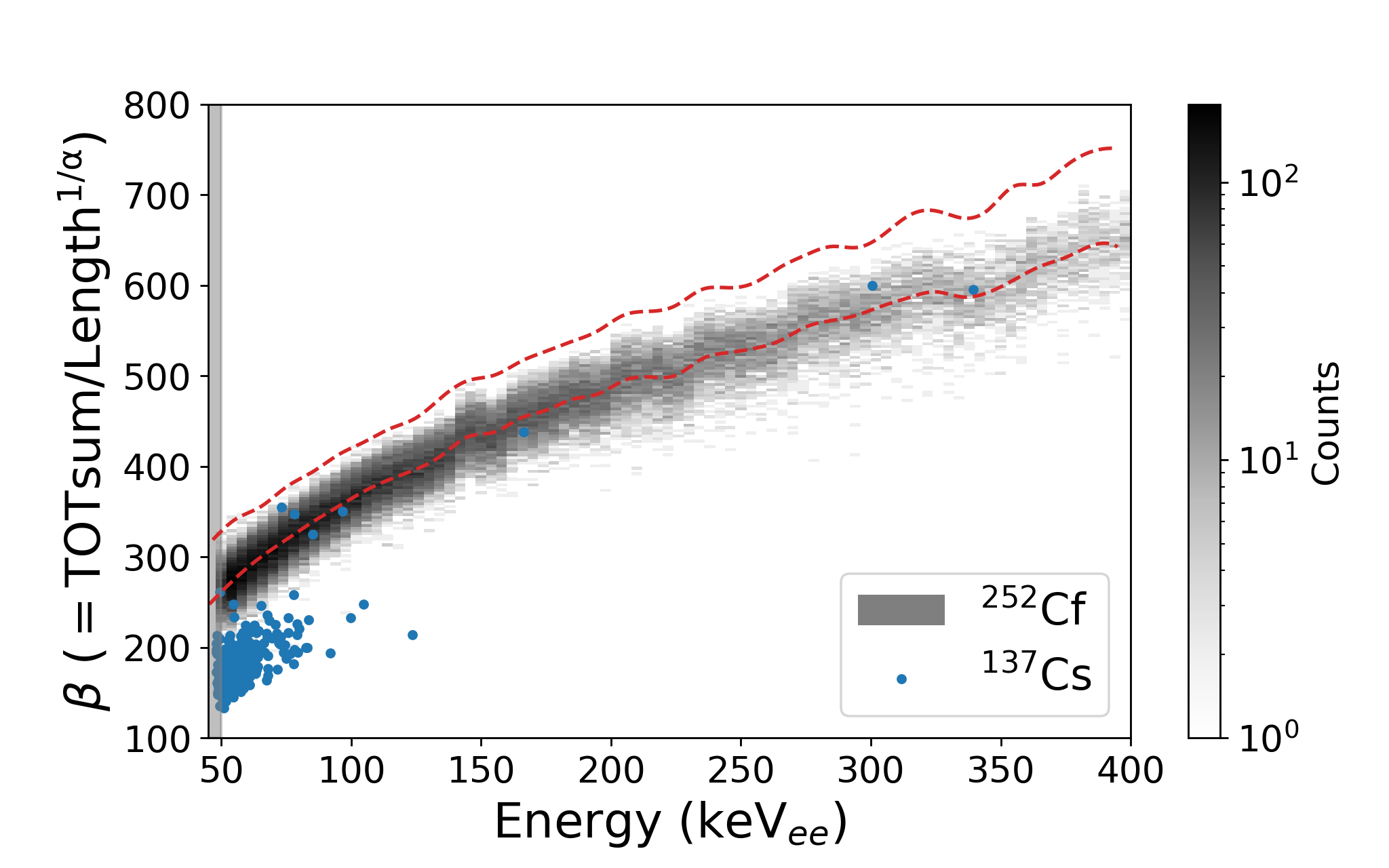}
    \caption{Energy dependence of $\beta$ at $\alpha$=2.3 after the TOTsum cut.
    The black gradient is for the $^{252}$Cf neutron source calibration data and the blue dots are for the $^{137}$Cs gamma-ray source calibration data.
    The dashed red lines indicate the mean  value and the 3$\sigma$ cut line by Gaussian fit, respectively.
    The events between the cut lines are selected.}
    \label{fig:beta-energy}
\end{figure}

\begin{figure}[h]
    \centering
    \includegraphics[width=0.8\textwidth]{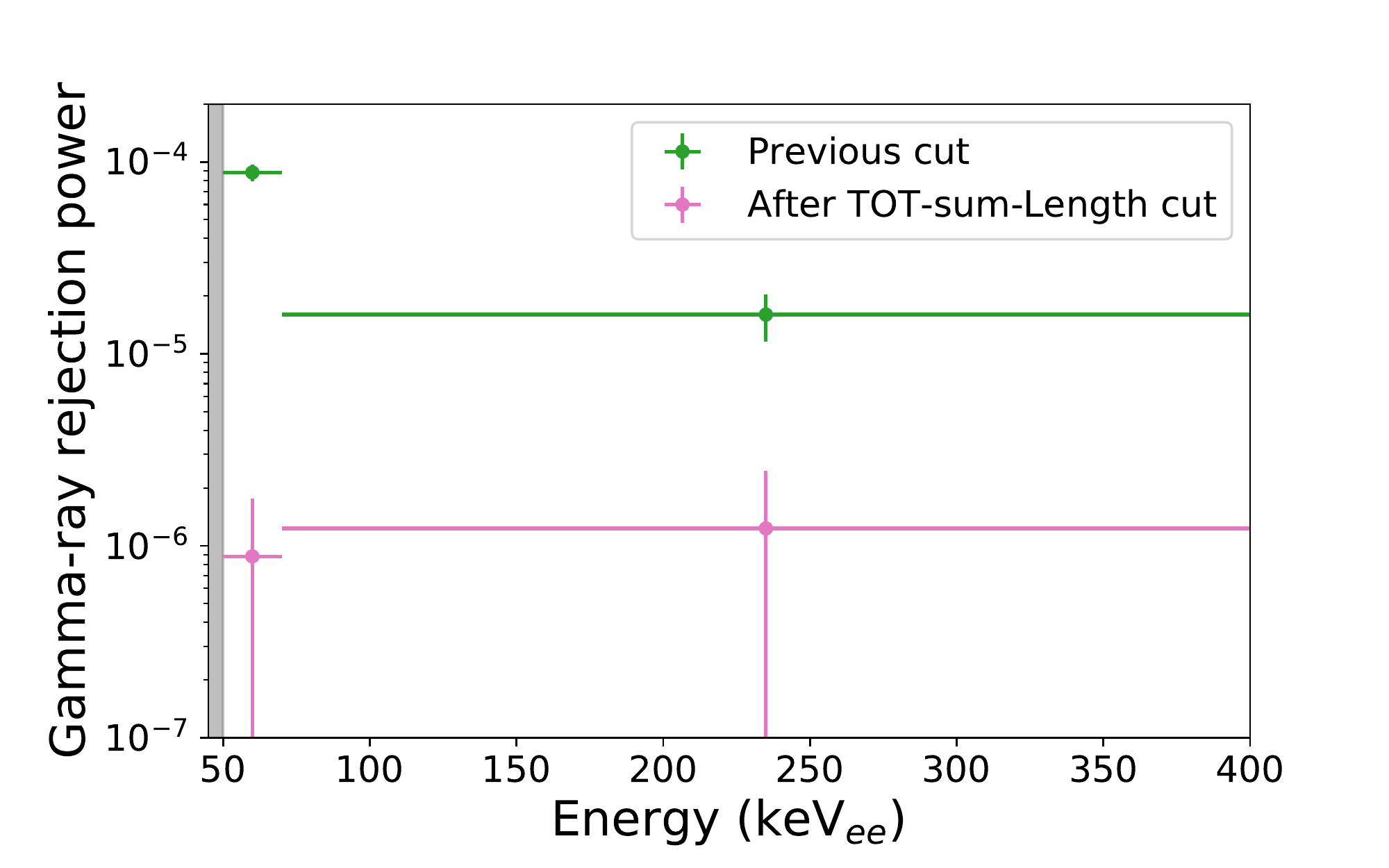}
    \caption{
    Gamma-ray rejection powers.
    The magenda dots are the result using the TOTsum-Length cut and green is the one without the TOTsum-Length cut.
    The TOTsum-Length cut introduced in this study improved the results by two orders of magnitude in the energy range of 50--70~keV$_{ee}$.}
    \label{fig:gamma-rej}
\end{figure}

We first determined $\alpha$ and $\beta$ values in the 50--60~keV$_{ee}$ energy range for each period.
The period is a set of data taken under a same detector condition and will be summarized in Section~\ref{sec:experiment}. 
The parameters were determined so that they would give the best rejection of gamma-ray events while retaining the selection efficiency of nuclear recoil events to be greater than 50\%.
Here, the selection efficiency for a specific selection is defined as the ratio of the remaining number of events to that before the selection.
We then fixed $\alpha$ and determined $\beta$ for a given energy.
Figure~\ref{fig:beta-energy} shows the energy dependence of $\beta$.
The black and blue dots represent the data with a $^{252}$Cf and a $^{137}$Cs sources, respectively.
The distribution of $\beta$ values of the nuclear recoils events was fit with Gaussian 
in every 10~keV energy bin.
The region between the mean and upper $3\sigma$ of the Gaussian indicated with red lines in Fig.~\ref{fig:beta-energy} was set as the nuclear recoil region and the rest was rejected.
Gamma-ray rejection powers with and without this cut are shown in Fig.~\ref{fig:gamma-rej}.
A gamma-ray rejection power of 8.8~$\times$~10$^{-7}$ was achieved, which is about two orders of magnitude better than that in NEWAGE2021.


\subsection{Head-tail analysis}
\label{sec:head-tail}
Importance of the track sense recognition, or the head-tail determination, has been stressed for years\,\cite{Anne_2007,Ciaran_2015}.
We started to use the head-tail determination for the direction-sensitive dark matter search analysis  
with a limited efficiency in Ref.~\cite{yakabe_ptep}.
An analysis update improved the efficiency 
and head-tail determinations for 3D tracks, or 3D-vector analysis, were used for this work.
The first step in reconstructing the direction of a track is to obtain the relative arrival times of ionized electrons in the readout strips.
These relative arrival times on X or Y strips are converted into relative Z positions taking account the drift velocity. 
The charge detected on the strip, or a hit, is thus assigned a (X, Z) or (Y, Z)  hit-position.
Angles of a track in the X-Z and Y-Z planes are known by fitting the hit-positions with straight lines.
3D-axial directions of the tracks in the detector coordinate system are determined from these two angles in the X-Z and Y-Z planes. 
These reconstructed tracks are not 3D-vector ones at this stage because the head-tail of the track is not determined yet.

The head-tail of a track can be determined
by observing the asymmetry of the energy deposition along its trajectory.
The fluorine-nuclear track with an energy of our interest (less than  400~keV$_{ee}$) is known to deposit its energy large at the starting point and small around its end point. 
This phenomena can be observed as 
large TOTs at the starting point and small TOTs around its end point.

\begin{figure}[h]
    \centering
    \includegraphics[width=1\textwidth]{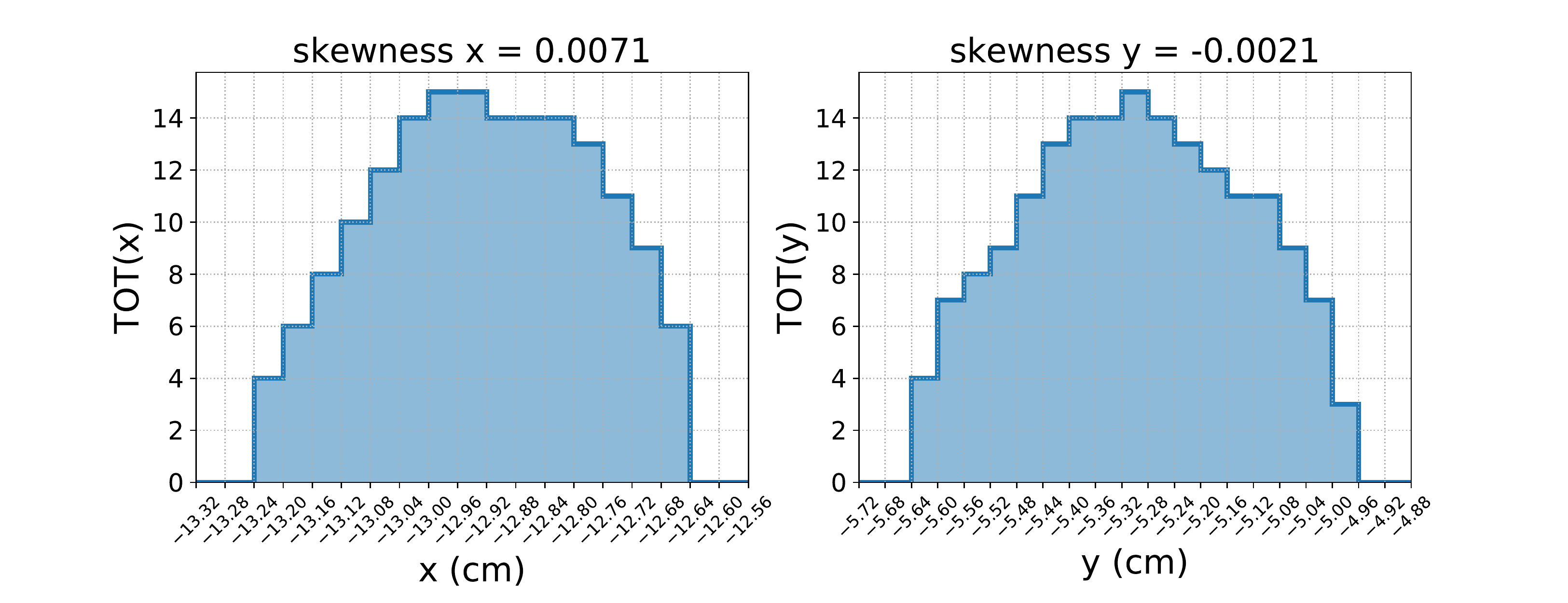}
    \caption{TOT values of an event along each X (left panel) and Y (right panel) strip.}
    \label{fig:tot-dist}
\end{figure}

Figure~\ref{fig:tot-dist} shows observed TOT distributions of an event along X and Y strips. This event was obtained with a $\rm{}^{252}Cf$ source placed at (25~cm, 0~cm, 0~cm) 
so that we expect to observe fluorine nucleus tracks running from +X to -X directions.
An asymmetry of the TOT distribution along the X-axis is seen while that along the Y-axis is more symmetric.
This asymmetry is quantified by parameters $skewneesses$ defined as following equations,
\begin{align}
    skewness~x & =  \frac{<TOT (x)\cdot (x-<x>)^3>}{< (TOT (x)\cdot (x-<x>)^2)^{3/2}>},\label{eq_skewnessX}\\ 
    skewness~y & =  \frac{<TOT (y)\cdot (y-<y>)^3>}{< (TOT (y)\cdot (y-<y>)^2)^{3/2}>}.\label{eq_skewnessY}
\end{align}
Here $TOT(x)$ is the TOT observed on strip $x$, and $< >$ represents the means value.  
The ability to determine the head-tail, called the head-tail power $P_{\rm ht}$, is defined as
\begin{equation}
    {P_{\rm ht}} = \frac{N_{\rm true}}{N},
    \label{eq:headtail}
\end{equation}
where $N$ is the total number of events, and $N_{\rm{true}}$ is the number of events head-tails of which were correctly determined by the skewness. Determinations of $N_{\rm{true}}$ are discussed later.

In our previous work, 
we selected events with small $\theta_{\rm ele}$ and 
large skewness to increase the head-tail power at a cost of 
lowering the selection efficiency to less than one half\,\cite{yakabe_ptep}.
The analysis was updated so that the the selection efficiency was recovered while the $P_{\rm ht}$ was retained; the use of $skewness~x$ and $skewness~y$ were determined according to the azimuth direction of the tracks.
For the tracks along the X-coordinate direction (0 $^{\circ} \leq$ $|\phi_{\rm{azi}}|$ $<$ 45 $^{\circ}$), $skewness~x$ 
was used, and $skewness~y$ was used for the tracks with 45$^{\circ} \leq$ $|\phi_{\rm{azi}}|$ $<$ 90$^{\circ}$). In addition, number of hit strips were increased by the operation at a high gas gains.

\begin{figure}[h]
    \centering
    \includegraphics[width=1\textwidth]{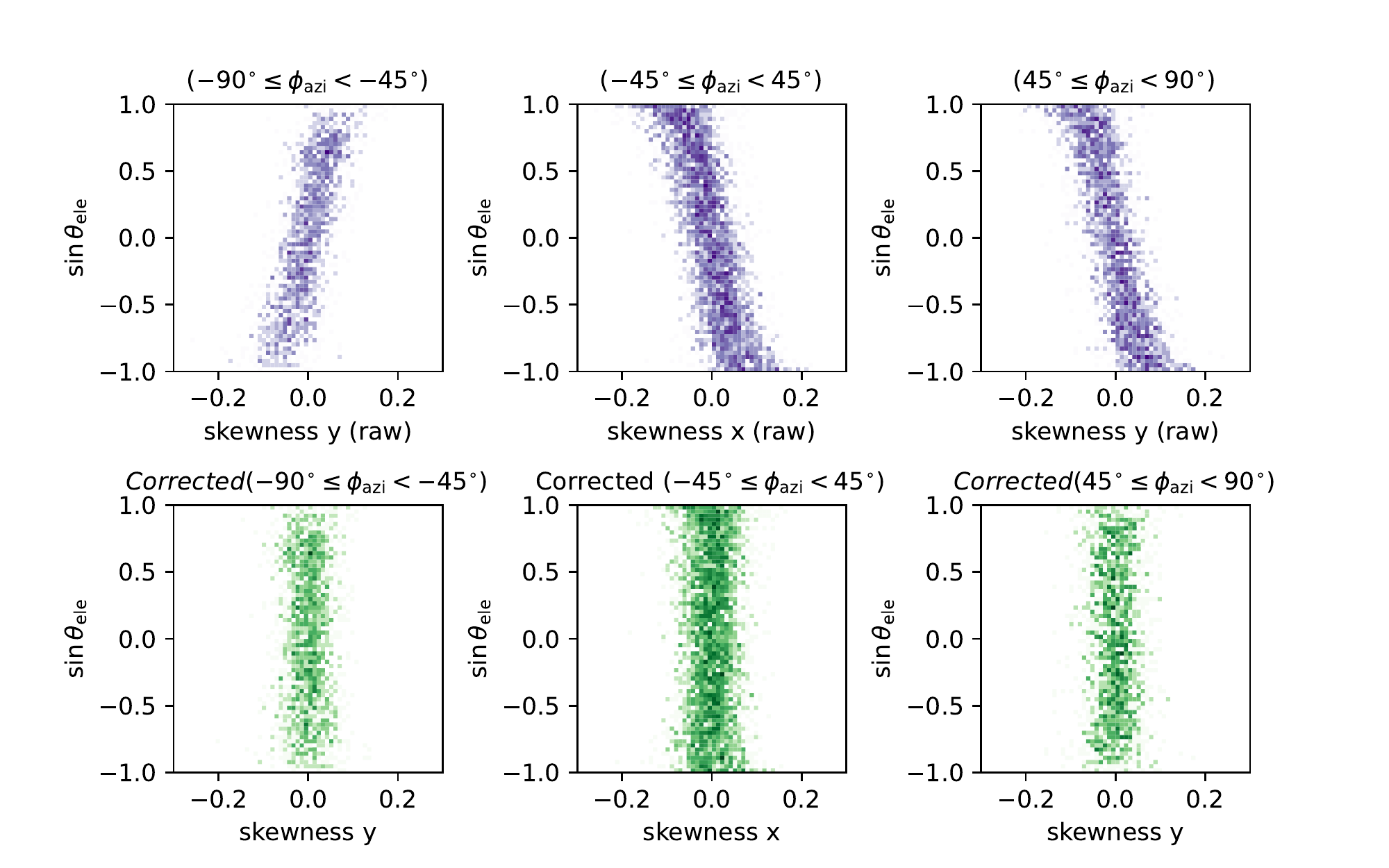}
    \caption{Correlation between skewnesses and $\sin\theta_{\rm{ele}}$ for the events in the energy range of 50--100~keV$_{ee}$. The distributions before and after the correction are shown in the upper and lower figures, respectively.}
    \label{fig:sin-skew}
\end{figure}

The original values of skewness were found to be correlated with $\theta_{\rm{ele}}$ in the measurement using $^{252}$Cf source. The upper panels of Fig.~\ref{fig:sin-skew} show the correlation between $\sin{\theta_{\rm{ele}}}$ and skewness in the $^{252}$Cf run. Here, since nuclear recoils scatter toward the direction of emitted neutrons from the $^{252}$Cf source, $\sin{\theta_{\rm{ele}}}$ was determined in a range of $[-1,1]$.
The skewness were corrected according to $\sin\theta_{\rm{ele}}$ with cubic functions, which are empirically decided,
and the corrected skewness values shown in the lower panels of Fig.~\ref{fig:sin-skew} were used for further discussions.

\begin{figure}[h]
    \centering
    \includegraphics[width=1\textwidth]{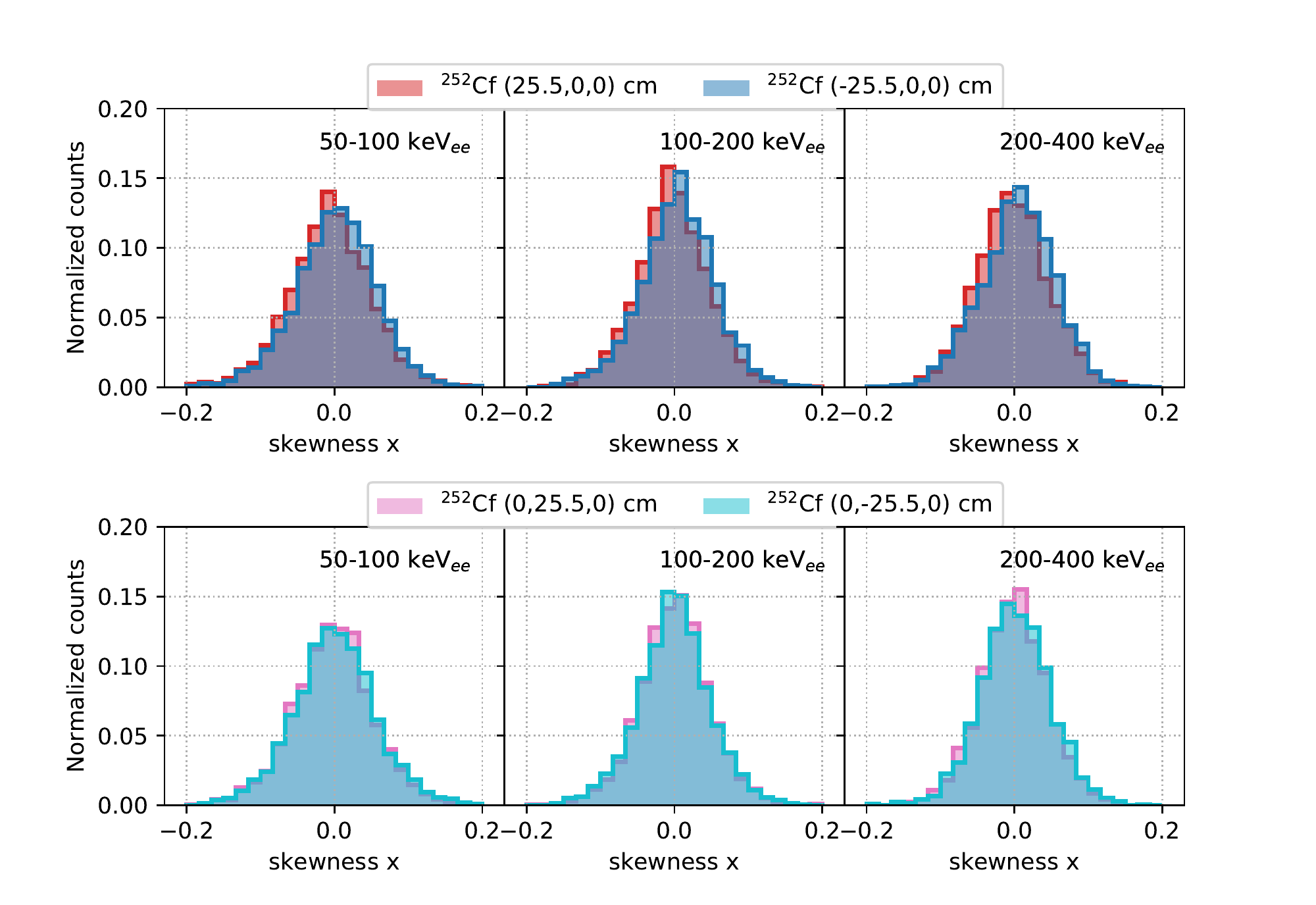}
    \caption{Distributions of $skewness~x$ for three energy ranges. In the top panel, the red and blue histograms show the neutron radiation data from +X and -X directions, respectively. The pink and cyan histograms in the bottom panel indicate the neutron radiation data from +Y and -Y directions, respectively.
    All histograms are normalized to unity.}
    \label{fig:skewx-dist}
\end{figure}

\begin{figure}[h]
    \centering
    \includegraphics[width=1\textwidth]{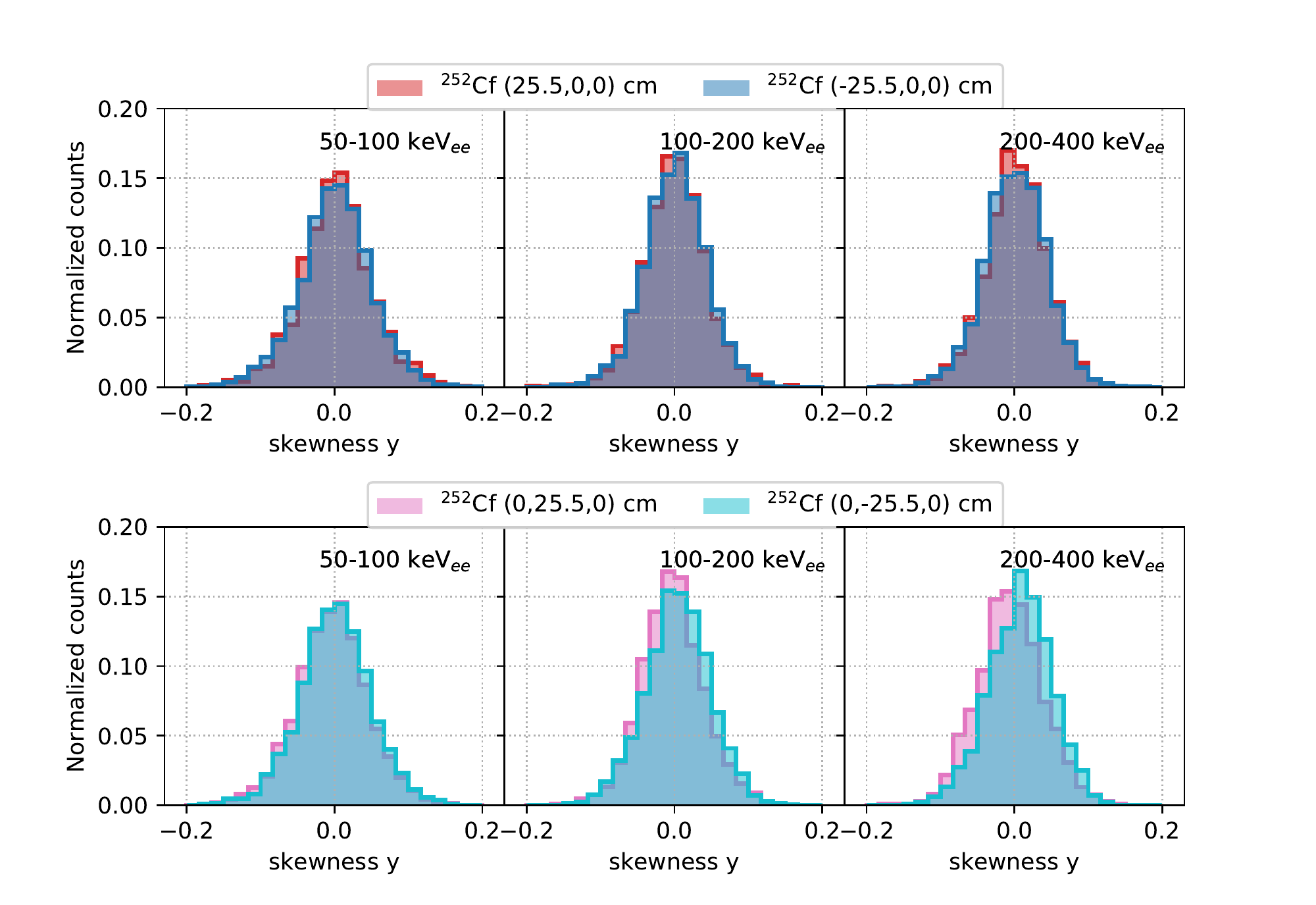}
    \caption{Distributions of $skewnes~y$ for three energy ranges. 
    In the top panel, the red and blue histograms show the neutron radiation data from +X and -X directions, respectively. The pink and cyan histograms in the bottom panel indicate the neutron radiation data from +Y and -Y directions, respectively.
    All histograms are normalized to unity.}
    \label{fig:skewy-dist}
\end{figure}

Figures~\ref{fig:skewx-dist} and~\ref{fig:skewy-dist} show skewness distributions 
of a $^{252}$Cf source data after all cuts for three energy ranges. Neutron irradiation data from $+$X and $-$X directions are shown with red and blue  histograms in the upper panels of Fig.~\ref{fig:skewx-dist}.
They show different $skewness~x$ distributions as expected while the $skewness~x$ distributions for the $\pm$Y direction irradiation data (lower panels of Fig.~\ref{fig:skewx-dist}) did not show significant difference. The same trend was confirmed for $skewness~y$ as shown in Fig.~\ref{fig:skewy-dist}. 
$N_{\rm true}$ was defined by discriminating at $skewness=0$. For instance, $N_{\rm true}$(+x) is defined as the number of events of $skewness~x<0$
of the red histograms in the top panel of Fig.~\ref{fig:skewx-dist}. On the other hand, $N_{\rm true}$(-x) is defined as the number of events of $skewness~x>0$ of the blue histogram. $P_{\rm ht}=50\%$ indicates that the detector has no sensitivity for the head-tail information.
Averaged $P_{\rm ht}$ values for 50--100~keV$_{ee}$, 100--200~keV$_{ee}$, and 200--400~keV$_{ee}$ energy ranges were $(52.4\pm1.1)\%$, $(52.9\pm1.4)\%$, and  $(53.6\pm2.0)\%$, respectively. 
Details of $P_{\rm ht}$ are summarized in Table~\ref{tab:headtail}.
The error of $P_{\rm ht}$ in each irradiation direction is the standard deviation of head-tail power determined for each period.
The overall head-tail power error is the standard deviation of the  $P_{\rm ht}$s 
in each irradiation direction.
Head-tail powers equivalent to those of Ref.\,\cite{yakabe_ptep} were achieved without any specific selection for the head-tail determination.

\begin{table}[htbp]
    \centering
    \begin{tabular}{cccccc} \hline\hline
        Energy range & $P_{\rm ht}$ (+x) (\%) & $P_{\rm ht}$ (-x) (\%)& $P_{\rm ht}$ (+y) (\%) & $P_{\rm ht}$ (-y) (\%) & $P_{\rm ht}$ (average) (\%) \\ \hline
        50--100~keV$_{ee}$ & 52.2$\pm$0.9 & 53.3 $\pm$1.2 & 52.2 $\pm$1.1 & 51.9 $\pm$0.9 & 52.4 $\pm$1.1 \\
        100--200~keV$_{ee}$ & 52.6 $\pm$1.4 & 53.2 $\pm$1.2 & 53.5 $\pm$1.2 & 52.5 $\pm$1.0 & 52.9 $\pm$1.2 \\
        200--400~keV$_{ee}$ & 53.3 $\pm$1.6 & 52.4 $\pm$1.0 & 54.9 $\pm$2.8 & 53.8 $\pm$1.6 & 53.6 $\pm$2.0 \\ \hline\hline
        \end{tabular}
    \caption{Head-tail powers in unit of \% for each direction and energy range.
    }
    \label{tab:headtail}
\end{table}

\subsection{Efficiencies}\label{sec:efficiency}
There are two types of efficiencies regarding this study;
the detection-selection and the directional efficiencies. 
The former, or the ``absolute'' efficiency, determines the number of detected-and-selected events while the latter, or the ``relative'' one, determines the directional distribution of these events without changing the total number of events. 
In order to determine the efficiencies including directionality, an isotropic data-set needs to be used.
The isotropic data-set was made by summing-up the time-normalized data obtained by irradiating the detector with neutrons from a $^{252}$Cf source placed at six positions in $\pm X$, $\pm Y$, and $\pm Z$ directions.

\begin{figure}[h]
    \centering
    \includegraphics[width=0.8\textwidth]{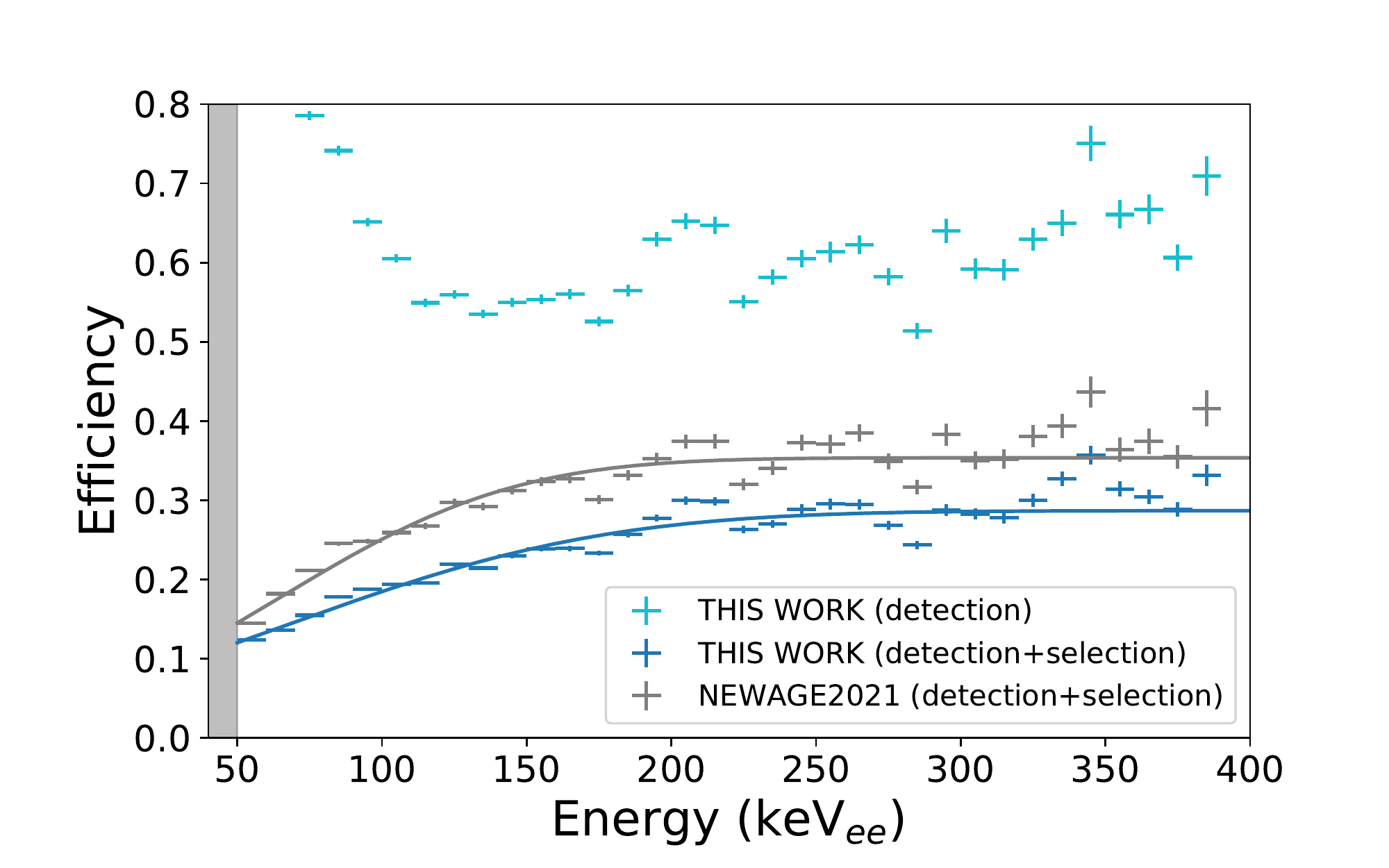}
    \caption{Nuclear recoil efficiencies as a function of the energy.
    The cyan and the blue histograms are the detection and detection-selection efficiencies of nuclear recoil of this study, respectively.
    The gray histograms is the result of NEWAGE2021
    \,\cite{ikeda_ptep}.}
    \label{fig:eff}
\end{figure}

The detection-selection efficiency is defined as the number of nuclear recoil events after all selections divided by the expected number of nuclear recoils
in the fiducial volume.
Here, the expected number of nuclear recoils is estimated by the Geant4 simulation.
Results are shown in Fig~\ref{fig:eff}.
It should be noted that the increase of the detection efficiency seen below 100~keV$_{ee}$ is due to the contamination of the 
gamma-ray events and is not real.
The contamination is removed with the selections to a negligible level.
The detection efficiency 
is about 60\% above 200~keV$_{ee}$.
The main reason 
of not reaching at 100\% 
is that the gas gain being not high enough to trigger all the nuclear recoil events.
The detection-selection efficiency above 200~keV$_{ee}$ is half of the detection efficiency
because of the mean value 
for the TOTsum-Length selection.
A 20\%-reduction of the 
detection-selection efficiency from NEWAGE2021 
should also attribute to the
additional cut, which still gives a large advantage in the signal-to-noise ratio if we consider the gain on the rejection shown in Fig.~\ref{fig:gamma-rej}.
The detection-selection efficiency 
shown in Fig.~\ref{fig:eff}, or the ''absolute'' efficiency, can be used to calculate the expected number of events for a given WIMP or background model. It can also be used to unfold the measured energy spectrum and obtain an ''effective'' spectrum for the comparison of the background rates.

\begin{figure}[h]
    \centering
    \includegraphics[width=0.8\textwidth]{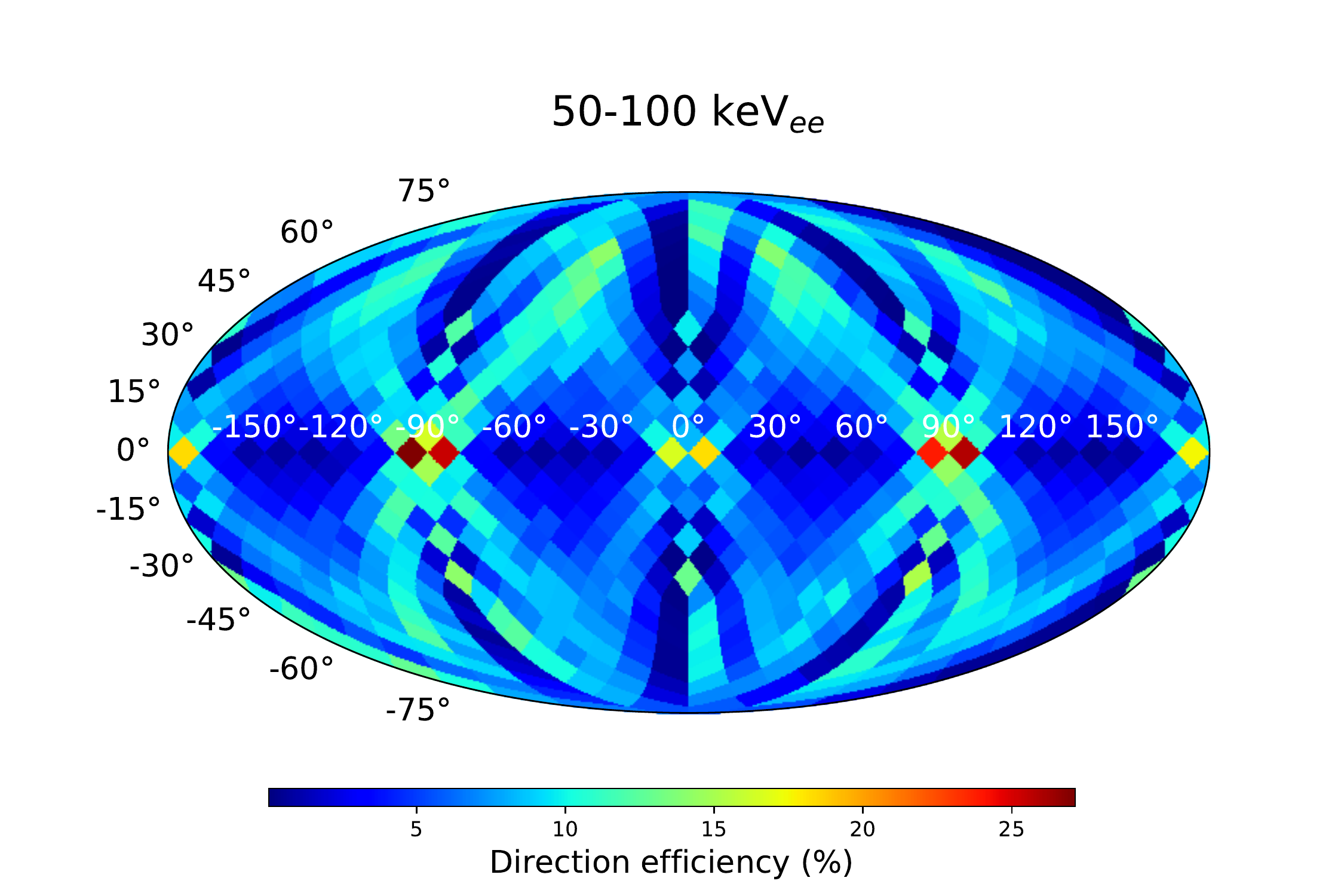}
    \caption{Directional efficiency in the detector coordinate system. The axes of the white and black labels are the azimuth angle $\phi_{\rm azi}$ and the elevation angle $\theta_{\rm ele}$, respectively. The color gradation indicates the relative reconstruction efficiency.}
    \label{fig:eff-dir}
\end{figure}

The directional efficiency is defined as the number of recoil events in a angular distribution, divided by the number of total recoil events.
Thus the directional efficiency is expressed as a sky map, or the relative response in the 
elevation ($\theta_{\rm{ele}}$) - azimuth ($\phi_{\rm{azi}}$) plane, for isotropic recoils.
The possible non-homogeneity of the directional efficiency mainly 
originates from the reconstruction algorithm. 
The 3D recoil direction, including the sense (head-tail) of the track, is reconstructed from the TOT-distributions of X and Y strips. 
Figure~\ref{fig:eff-dir} shows the obtained $\theta_{\rm{ele}}$-$\phi_{\rm{azi}}$ distributions of an isotropic recoil calibration data. Since this map is to know the ''relative'' or reconstruction efficiency of the directions,  the color map is a relative one to be used with the total number of events being conserved.
It is seen that 
the tracks tend to be reconstructed to align with the strips, $i. e.$ $\phi_{\rm azi}=0^{\circ}, \pm 90^{\circ}, 180^{\circ}$
for the tracks parallel to the detection plane, or the tracks with $\theta_{\rm ele}\sim0 $. 
The directional efficiencies shown in Fig.~\ref{fig:eff-dir}, or the relative efficiency, can be used to make an expected recoil distribution for a given number of expected events calculated by the detection-selection efficiency. 



\subsection{Angular resolution}\label{sec:angular-resolution}
The angular resolution was evaluated by comparing the distribution of the recoil angle $\gamma$ of neutron irradiation data with the simulated ones smeared by various angular resolutions.
Here $\gamma$ is the angle between the incoming neutron direction and the reconstructed nuclear-recoil direction.
Since the head-tails of the tracks are determined and considered in the analysis independent from the effect of the angular resolution, 
the angular resolution was evaluated with the distribution of absolute value of $\cos\gamma$.
$\chi^2_{\rm{ang}}$ value defined by 
Eq.~(\ref{eq:chi-ang}) was calculated for a given angular resolution $\sigma_{\rm{ang}}$.
\begin{align}
    \chi^2_{\rm{ang}} = \sum_i^{N_{\rm{bin}}} \frac{(N^{\rm{data}}_{i}- N^{\rm{MC}}_{i}(\sigma_{\rm{ang}}))^2}{N^{\rm{data}}_{i}},
    \label{eq:chi-ang}
\end{align}
where $N^{\rm{data}}_{i}$ is the number of events in the $i$-th bin of the histogram of measured $|\cos\gamma|$, and $N^{\rm{MC}}_{i}$ is the number of events in the $i$-th bin of the histogram of the $|\cos\gamma|$ distribution simulated by Geant4 smeared with the angular resolution, and $N_{\rm{bin}}$ is the number of bins in that histogram.
The angular resolution at the minimum $\chi^2_{\rm{ang}}$ value was adopted. 
The angular resolution was 58.1$_{-2.8}^{+5.8}$~degree in the energy range of 50--100~keV$_{ee}$.

\section{Experiment}
\label{sec:experiment}
A direction-sensitive dark matter search was performed in Laboratory B, Kamioka Observatory (36.25'N, 137.18'E), located 2700~m water equivalent underground.
The measurement was carried out from December 12th, 2017 to March 26th, 2020, subdivided into eight periods. 
The period was renewed when the detector was evacuated and filled with new CF$_{4}$ gas. The period information is summarized in Table~\ref{tab:run_summary}.
The Z-axis of the NEWAGE-0.3b'' detector was aligned to the direction of S30$^{\circ}$E.
The target gas was CF$_4$ at 76~Torr (0.1~atm) with a mass of 10~g in an effective volume of 28~$\times$~24~$\times$~41~cm$^3$ (27.6~L).
The total live time is 318~days corresponding to an exposure of 3.18~kg$\cdot$days\higashino{, which is 2.4 times larger exposure than that of NEWAGE2021}.

\begin{figure}[h]
    \centering
    \includegraphics[width=1.0\textwidth]{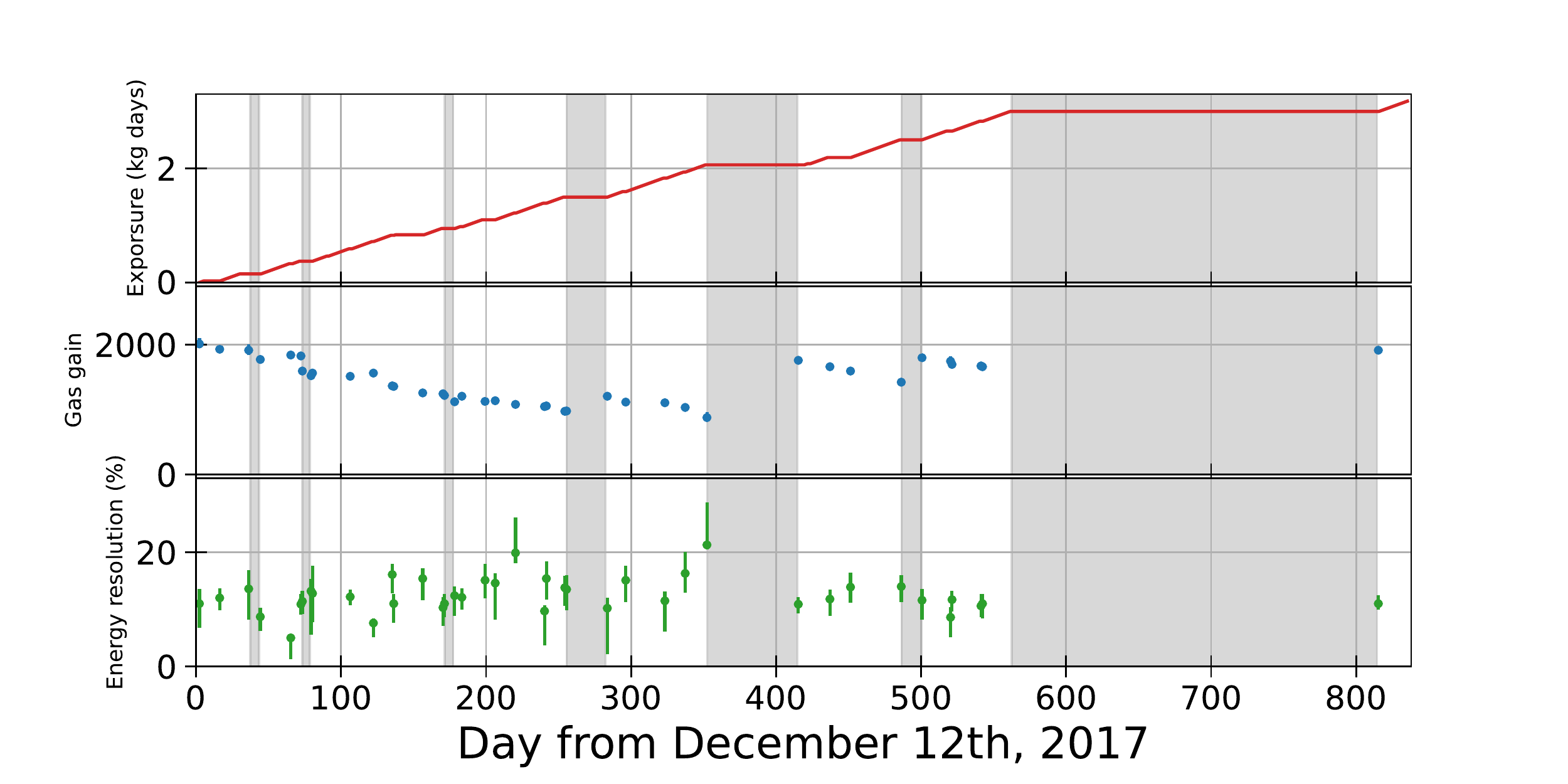}
    \caption{Cumulative exposure, gas gains, and energy resolutions during the measurement.
    }
    \label{fig:run_monitor}
\end{figure}

Various environmental parameters were monitored during the measurement to confirm the stability of the detector.
Figure~\ref{fig:run_monitor} shows the time dependences of the integrated exposure, the gas gain and the energy resolution. 
The energy calibrations and the efficiency measurements were performed approximately every two weeks.
The energy scale was corrected by the monitored gas gain.
The mean value of the energy resolution was 12.4\% with a standard deviation of 3.0\% during the measurement.
No variation of the energy resolution beyond errors was observed.

\begin{table}[htbp]
    \centering
    \begin{tabular}{ccccc}\hline\hline
        Period & Date & Gas gain & Live time (days) & Exposure (kg$\cdot$days) \\ \hline
        RUN20-1 & 2017/12/12 -- 2018/01/18 & 2000 & 13.5  & 0.135 \\
        RUN20-2 & 2018/01/23 -- 2018/02/23 & 1750 & 20.0  & 0.200 \\
        RUN21   & 2018/02/28 -- 2018/06/01 & 1550 & 58.6  & 0.586 \\
        RUN22-1 & 2018/06/06 -- 2018/08/24 & 1110 & 52.5  & 0.525 \\
        RUN22-2 & 2018/09/20 -- 2018/11/29 & 1200 & 60.5  & 0.605 \\
        RUN23   & 2018/12/05 -- 2019/04/12 & 1750 & 45.9  & 0.459 \\
        RUN24   & 2019/04/26 -- 2019/06/27 & 1800 & 49.4  & 0.494 \\
        RUN25   & 2020/03/04 -- 2020/03/26 & 1950 & 17.6  & 0.176 \\ \hline
        Total   & 2017/12/12 -- 2020/03/26 &      & 318.0 & 3.180 \\
        \hline\hline
    \end{tabular}
    \caption{Summary of the measurement periods with gas gains (at the start of each RUN), live times, and exposures. RUN22-1 and RUN22-2 are the data analyzed in NEWAGE2021\,\cite{ikeda_ptep}.}
    \label{tab:run_summary}
\end{table}

\begin{figure}[h!]
    \centering
    \includegraphics[width=0.8\textwidth]{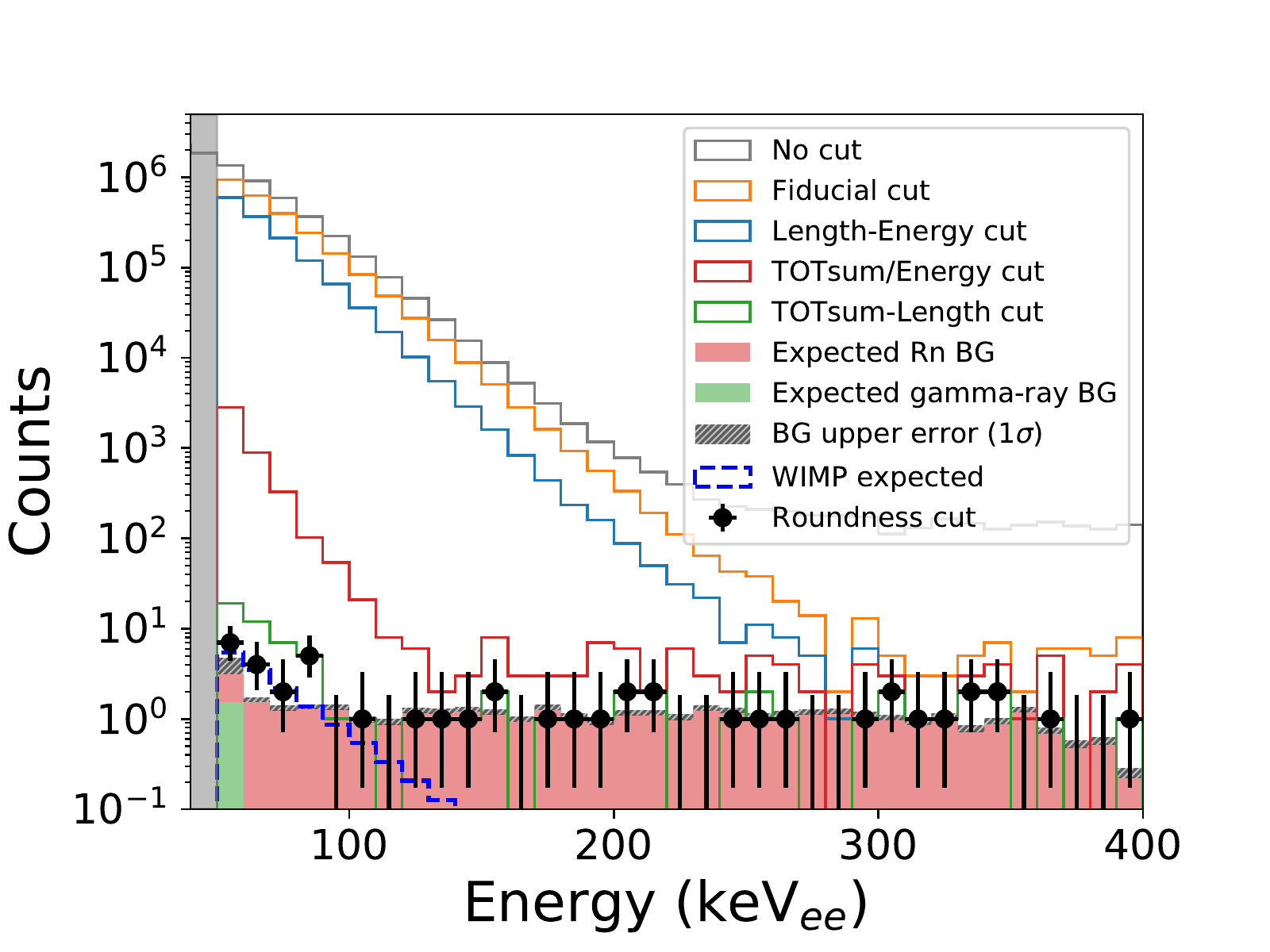}
    \caption{
    Energy spectra after each selection step.
    The grey, orange, blue, magenta, and green lines are the energy spectra after no cut, Fiducial volume cut, Length-Energy cut, TOTsum/Energy cut, and TOTsum-Length cut, respectively.
    The black dots with error bars are the final data sample after the Roundness cut.
    The fill stacked green and red spectra are the expected gamma-ray and radon background ones estimated by the simulation. The gray shaded area is a $1\sigma$ error in the background.
    The dashed blue line shows the expected spectrum by WIMPs assuming the mass of 150~GeV and the cross-section of 14.3~pb.
    }
    \label{fig:spectrum}
\end{figure}

The event selections described in subsection~\ref{sec:conv-selection} and \ref{sec:new-selection} were applied to the data.
Figure~\ref{fig:spectrum} shows the energy spectrum after each event selection.
The statistical errors are shown for the spectrum after all selections.
We confirm that the background in this work and NEWAGE2021 was consistent.

\begin{figure}[h]
    \centering
    \subfloat[Nuclear-recoil directions in the detector coordinate]{
        \includegraphics[width=0.7\textwidth]{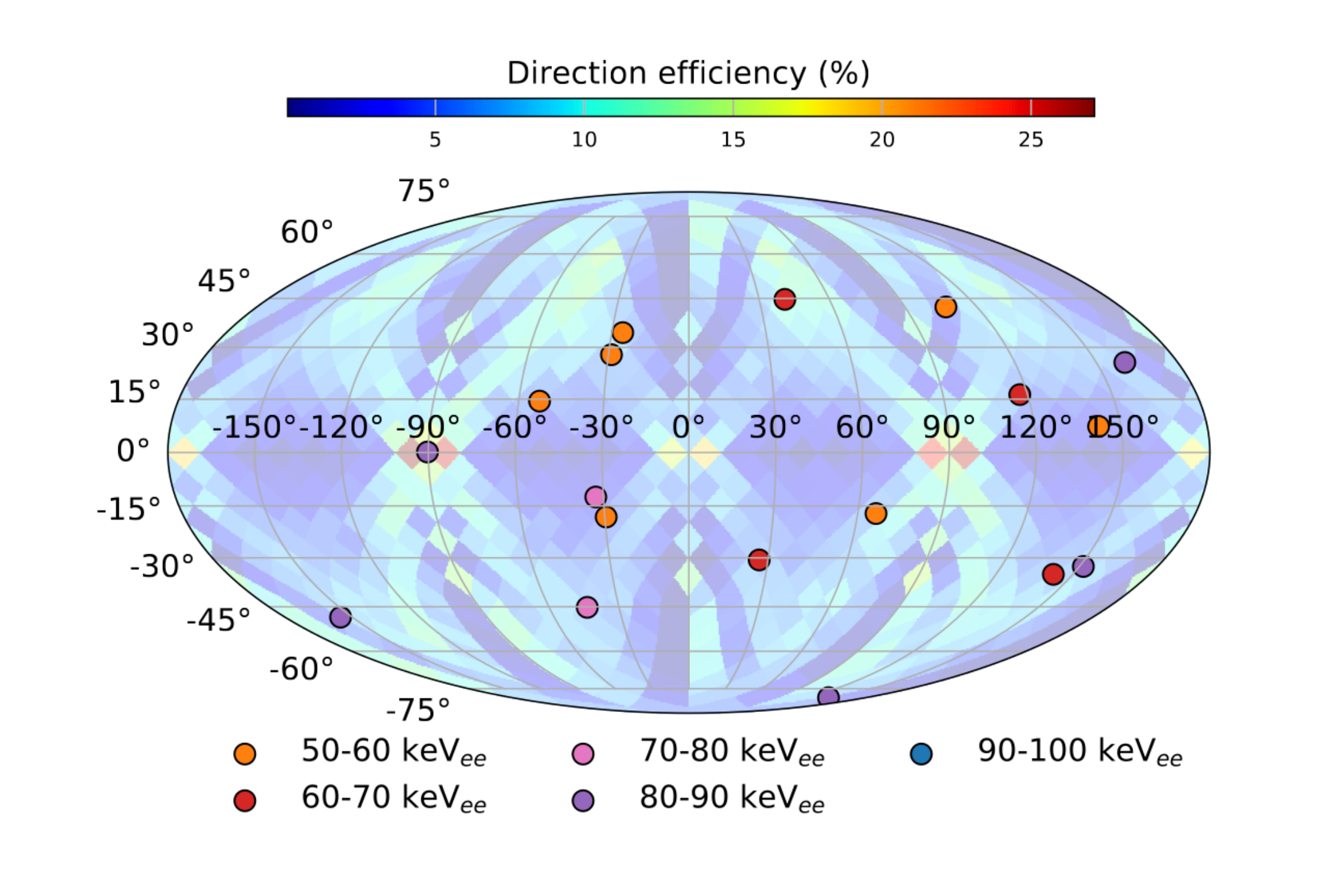}
    }\\
    \subfloat[Nuclear-recoil directions in the galaxy coordinate]{
        \includegraphics[width=0.7\textwidth]{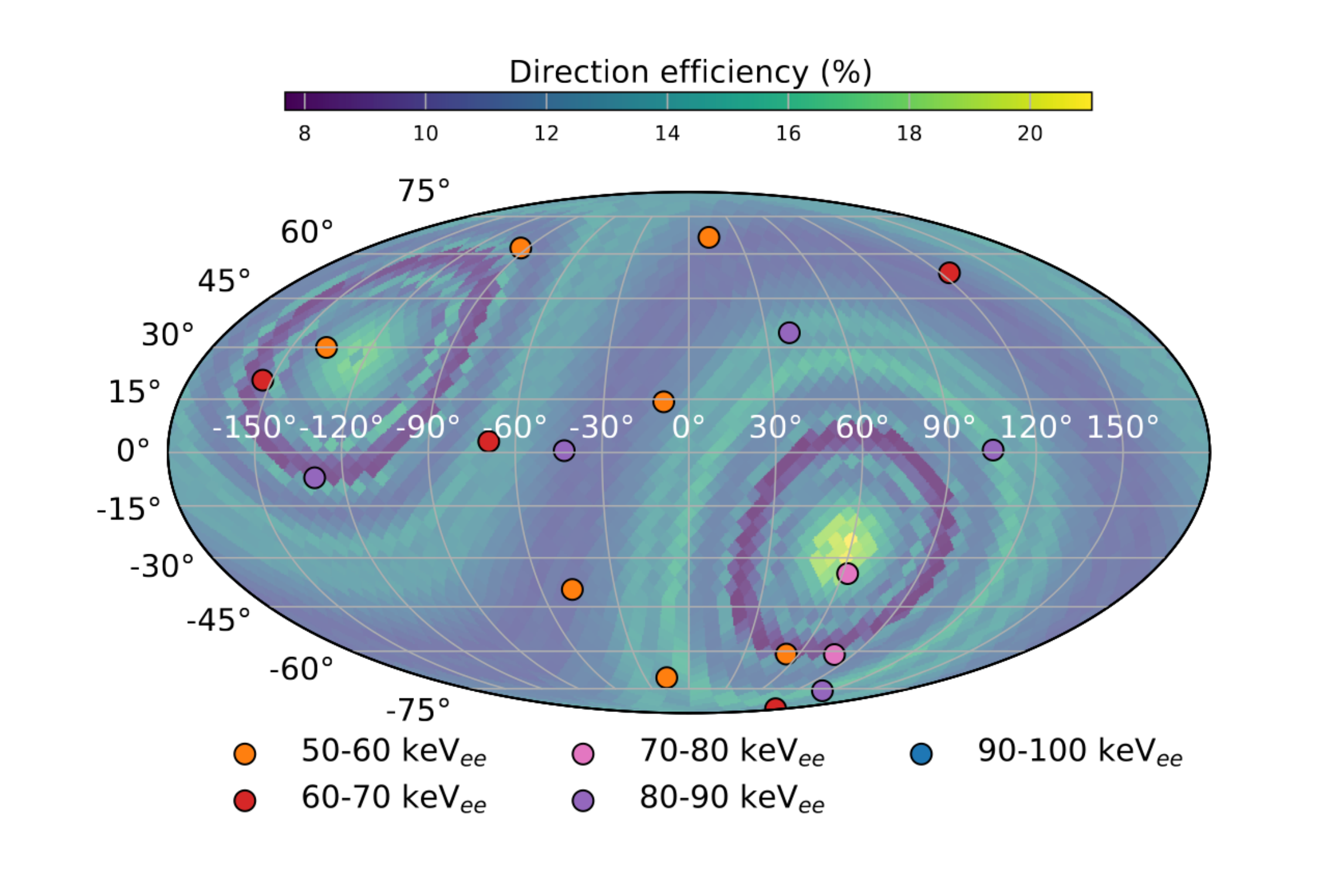}
    }
    \caption{(a) Nuclear recoil directions of final data sample in the detector coordinate. The X-axis and Y-axis are $\phi_{\rm{azi}}$ and $\theta_{\rm{ele}}$ in the detector coordinate system, respectively.
    (b) Nuclear recoil directions of final data sample in the galactic coordinate.
    The X-axis and Y-axis are the longitude and latitude of the galactic coordinate, respectively.
    The direction of the galactic center is (0,0) and that of Cygnus is (-90,0).
    The orange, red, pink, purple, and blue points indicate the energy ranges of 50--60~keV$_{ee}$, 60--70~keV$_{ee}$, 70--80~keV$_{ee}$, 80--90~keV$_{ee}$, and 90--100~keV$_{ee}$, respectively.
    The color contours in the background are the directional efficiencies in each coordinate system.}
    \label{fig:skymap}
\end{figure}

Figure~\ref{fig:skymap} shows the directions of measured nuclear recoil events in the detector coordinate (a) and the galactic coordinate (b), respectively. 
The $\cos\theta_{\rm{CYGNUS}}$ was calculated for each event in Fig.~\ref{fig:skymap} (b) and distributions are shown in Fig.~\ref{fig:fit-cos-dist}.
The $\cos\theta_{\rm{CYGNUS}}$ is binned by four and the energy is binned every 10~keV$_{ee}$.

\section{Results}
\label{sec:result}
\begin{figure}[!h]
    \centering
    \includegraphics[width=1.\textwidth]{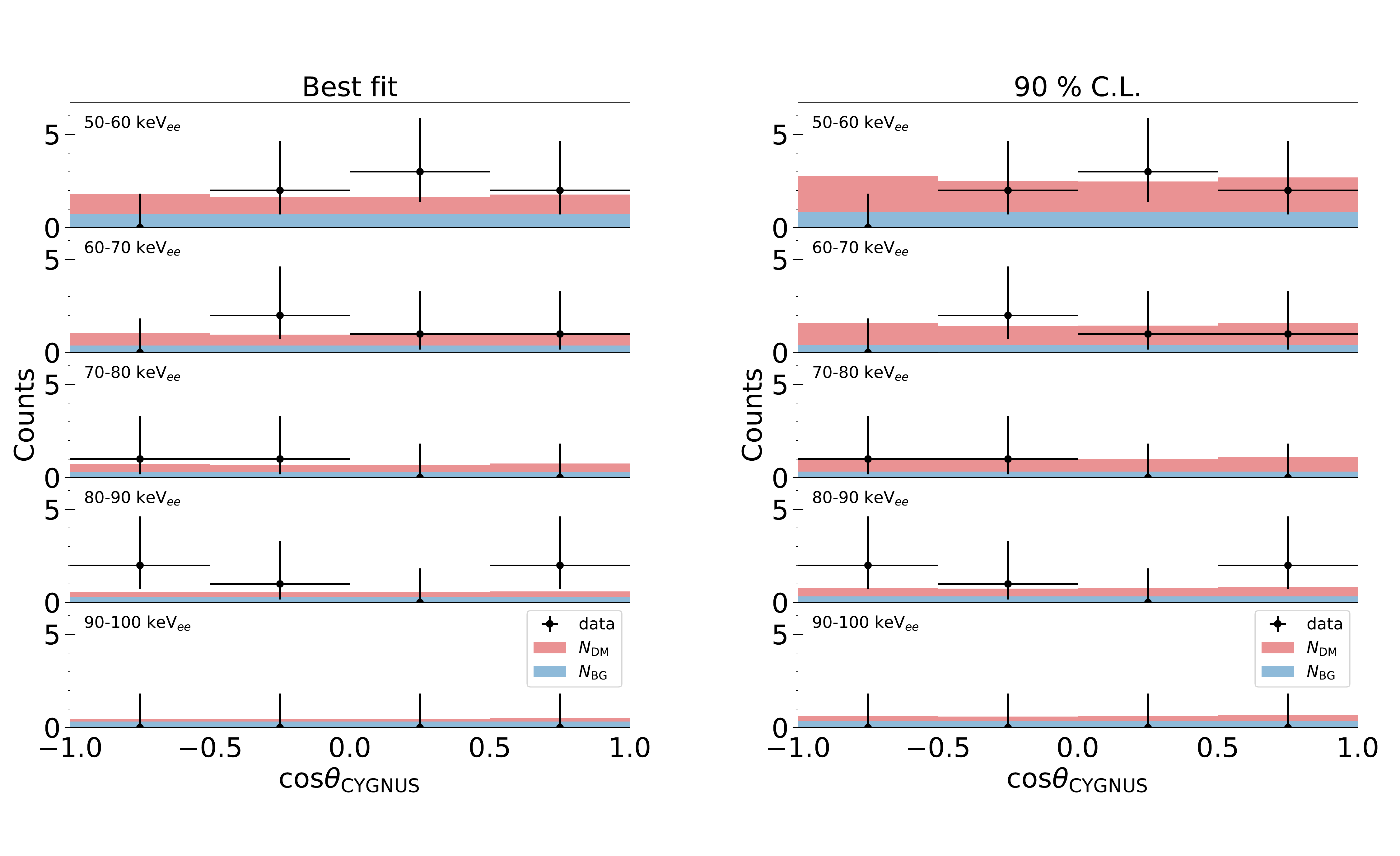}
       \caption{$\cos\theta_{\rm{CYGNUS}}$ distributions (identical black histograms in both panels) for the final date sample in the 50--100~keV$_{ee}$ energy ranges. The best fit and 90\% upper limit distributions for the WIMP mass of 150~GeV/$c^{2}$ are shown with color histograms in the left and right panels, respectively.}
    \label{fig:fit-cos-dist}
\end{figure}

A directional WIMP search analysis was performed with an assumption of the standard halo model.
Here the Maxwell distribution with a velocity dispersion of 220~km/sec, and an escape velocity of 650~km/sec were assumed\,\cite{PIFFL}. 
The local density of 0.3~$\rm GeV / c^2/cm^3$ was assumed. The spin parameter $\lambda^2 J(J+1)$ for the $\rm {^{19}F}$ of 0.647 was used in this analysis\, \cite{LEWIN199687}.
Considering the nuclear quenching factor, the simulated energy spectrum by the WIMP-nuclear scattering was re-scaled by SRIM~\cite{SRIM}, which represented the observed alpha-ray in the previous experiment~\cite{Nishimura2009}.
The spectra of $\cos\theta_{\rm{CYGNUS}}$ for each energy bin as shown in Fig.~\ref{fig:fit-cos-dist} 
were simultaneously compared with sum distributions of the WIMP signal and isotropic background using the binned likelihood ratio method.

A statistic value $\chi^2$ was defined as Eq.~(\ref{eq_X2}).
\begin{align}
    \chi^{2} &= 2\sum_{i=0}^n \sum_{j=0}^m \biggl[ ( N_{i,j}^{\rm{MC}}-N_{i,j}^{\rm{data}} ) + N_{i,j}^{\rm{data}}{\rm ln}\biggl (\frac{ N_{i,j}^{\rm{data}}}{N_{i,j}^{\rm{MC}}}\biggr) \biggr] + \alpha_{\rm{E}}^{2} + \alpha_{\rm{BG}}^{2}, \label{eq_X2}
\end{align}
where,
\begin{align}
    N^{\rm{MC}}_{i,j} &= N_{i,j}^{\rm{DM}} (\sigma_{\chi-p}, m_{\chi}, \xi_{\rm{E}}) + N_{i,j}^{\rm{BG}} (\xi_{\rm{E}}, \xi_{\rm{BG}}), \label{eq_Nexp}\\
    \alpha_{\rm{E}} &= \frac{\xi_{\rm{E}}}{\sigma_{\rm{E}}}, \label{eq_alphaE}\\
    \alpha_{\rm{BG}} &= \frac{\xi_{\rm{BG}}}{\sigma_{\rm{BG}}} \label{eq_alphaBG}.
\end{align}
Subscripts $i$ and $j$ are the bin-number of the $\cos\theta_{\rm{CYGNUS}}$ and the energy, respectively.
The expected and measured number of events in bin $i,j$ are described as $N_{i,j}^{\rm{MC}}$ and $N_{i,j}^{\rm{data}}$,     respectively.
$N_{i,j}^{\rm{MC}}$ is written as Eq.~\ref{eq_Nexp}, where 
$N_{i,j}^{\rm{DM}}$ is the expected number of the WIMP-nucleus scatterings,
and $N_{i,j}^{\rm{BG}}$ is the expected number of background events.
$\sigma_{\chi-p}$ is the WIMP-proton cross section.
$N_{i,j}^{\rm{BG}}$ was estimated using the Geant4 simulation based on the flux measurements of the ambient gamma-rays, the ambient neutrons, the alpha rays from the radon, and the alpha rays from the LA$\mu$-PIC surface.
The dominant background components in the energy range of 50--100~keV$_{ee}$ were the ambient gamma-rays and the alpha rays from the radon (see Ref.\,\cite{ikeda_ptep} for details).
Expected background spectra are shown in Fig.~\ref{fig:spectrum} for reference.
The largest systematic uncertainty of the expected rate arise from the energy scale uncertainty.
This uncertainty was estimated from the discrepancy of the energy calibration between $^{10}$B, $^{220}$Rn, and $^{222}$Rn measurements discussed in subsection~\ref{sec:calibration}.
The uncertainty was evaluated in each run.
The weighted average of the energy scale uncertainty was +13.2\% and -2.3\%.
The uncertainties of the background rate are the measurement errors of radioactivities for the ambient gamma-rays and the radons.
Here the ambient gamma-ray flux was measured with a CsI scintillator~\cite{NishimuraPhD} and the radon background was estimated with the high energy spectrum of this work. 
Nuisance parameters $\alpha_{\rm{E}}$ and $\alpha_{\rm{BG}}$ considering the systematic uncertainty from the energy scale $\sigma_{\rm{E}}$ and the background estimation $\sigma_{\rm{BG}}$ are defined as Equations (\ref{eq_alphaE}) and (\ref{eq_alphaBG}).
Possible shifts of the energy scale and the number of expected backgrounds are expressed as $\xi_{\rm{E}}$ and $\xi_{\rm{BG}}$. 
$\chi^2$ was minimized 
for a given WIMP mass with 
$\sigma_{\chi-p}$, pull-terms $\alpha_{\rm{E}}$ and $\alpha_{\rm{BG}}$ 
as fitting parameters. 
We first explain the procedure for the WIMP mass of 150~GeV/$c^{2}$ case.
A minimum $\chi^2$/NDF of 20.4/17 was obtained for $\sigma_{\chi-p}$=14.6~pb.
The left panel in Fig.~\ref{fig:fit-cos-dist} shows the 
$\cos\theta_{\rm{CYGNUS}}$ distributions of the best-fit case.
A chi-square distribution was created from a dummy sample of isotropic background model using Monte Carlo simulations. 
This test gave the p-value of 0.60 for the measured result.
Observed distribution was thus found to be consistent with the background-only model. 
Since no significant WIMP excess  was obtained, an upper limit at 90\% confidence level (C.L.) was set for the spin-dependent WIMP-proton scattering cross section.
The likelihood ratio $\mathcal{L}$ is defined as,
\begin{equation}
    \mathcal{L}={\rm{exp}}\biggl (-\frac{\chi^{2} (\sigma_{\chi-p})-\chi^{2}_{\rm min}}{2} \biggr).
\end{equation}
Here, $\chi^{2} (\sigma_{\chi-p})$ and $\chi^{2}_{\rm min}$ are the value of $\chi^{2}$ and the minimum value of $\chi^{2}$ calculated by varying $\sigma_{\chi-p}$, respectively.
The 90\% C.L. upper limit of the WIMP-proton cross section, $\sigma_{\chi-p}^{\rm limit}$ , is determined as follows,
\begin{equation}
    \frac{\int_{0}^{\sigma_{\chi-p}^{\rm{limit}}} \mathcal{L} d\sigma_{\chi-p}} {\int_{0}^{\infty} \mathcal{L} d\sigma_{\chi-p}} = 0.9.
    \label{eq:feldman}
\end{equation}
Using the above equation, the 90\% C.L. upper limit of the spin-dependent cross section was found to be 25.7~pb for a WIMP mass of 150~GeV/$c^{2}$.
The $\cos\theta_{\rm{CYGNUS}}$ distributions with  
the upper limit of 90\% C.L. are shown in the right panels of Fig.~\ref{fig:fit-cos-dist}.

\begin{figure}[h]
    \centering
    \includegraphics[width=1.0\textwidth]{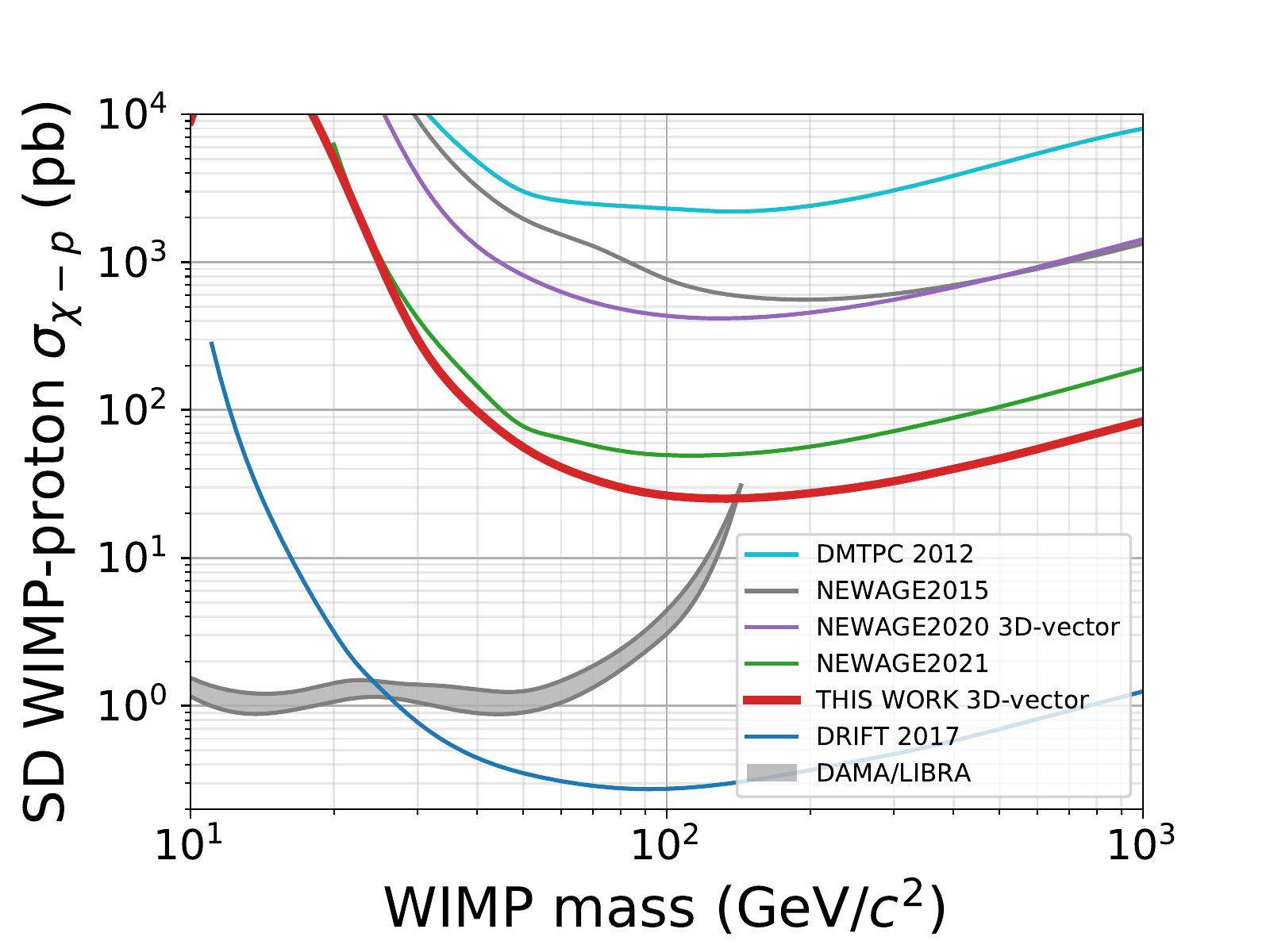}
    \caption{90\% C.L. upper limits of the spin-dependent WIMP-proton scattering cross section as a function of the WIMP mass.
    The red line is the result of this work.
    The limit calculated without the head-tail information was almost identical to this work.
    The green line is the result of 
    our previous work (NEWAGE2021\,\cite{ikeda_ptep}) and the purple line is the result with 
    the 3D-vector directional analysis for NEWAGE2020\,\cite{yakabe_ptep}.
    The gray line is the result of NEWAGE2015\,\cite{nakamura_ptep}.
    The solid light-blue shows the results from the directional analysis of DMTPC\,\cite{DMTPC}.
    The blue line is the limit curve for DRIFT\,\cite{DRIFT2017}.
    It should be noted that the upper limit of the DRIFT 2017 is led from the only energy information.
    The gray area is an interpretation of the allowed region of DAMA/LIBRA\,\cite{DAMA}.}
    \label{fig:limit}
\end{figure}

Upper limits of the cross sections were obtained for other WIMP masses by the same procedure.
Figure~\ref{fig:limit} shows the upper limits at 90\% C.L. of the spin-dependent WIMP-proton cross sections as a function of the WIMP mass.
Compared to the NEWAGE2020 results, which was analyzed by the 3D-vector method using the standard $\mu$-PIC, this upper limit updates by about one order of magnitude.
This is due to the reduction of surface background events with the LA$\mu$-PIC.
Furthermore, compared to the NEWAGE2021 result, the statistics of the 2.4 factor and an updated analysis including the background estimation, improved the limits by a factor of about two for WIMPs heavier than 100~GeV/$c^{2}$.
In the NEWAGE2021 analysis, the exclusion limit in the WIMP mass below 100~GeV/$c^{2}$ was better than expected due to the statistical fluctuation. This work improved the statistical uncertainty, in addition to the improvement of background estimation. This results in the update of the exclusion limit curve in any WIMP mass region.
We marked the most stringent limit via the directional analysis.

The 3D-vector method was successfully implemented in the analysis without impact on the upper limit of the WIMP-proton cross section in the background-dominant analysis.
The directional analysis including head-tail information demonstrates the possibility to reveal the property of WIMP.



\section{Discussions}
A new limit by a directional dark matter search with a 3D-vector analysis was obtained by this work. 
Although we started to search the region of one of the interpretations of the 
DAMA/LIBRA's annual modulation signal\,\cite{DAMA}, a significant improvement of the sensitivity is needed for the search of the region of more interest. 
The improvements can be realized mainly in three aspects: the detection-selection efficiency, the energy threshold, and the backgrounds.

The detection-selection efficiency at 50--60~keV$_{ee}$ is 12.5\%, which indicates the statistics can be increased by a factor of eight at most for a same exposure by an improvement of the detection-selection efficiency.
A measurement with a higher gas gain will increase the trigger efficiency.
A better gamma-ray rejection analysis, $e. g.$ introducing machine-learning methods, would compensate the expected increase of the gamma-ray background rate and allow us to operate the detector at a higher gas gain. Shielding the detector is an independent hardware approach to reduce the gamma-ray background events.

The current energy threshold (50~keV$_{ee}$) is mainly limited by the track length of the recoil events.
Typical length of the track of fluorine nuclear recoil below 50~keV$_{ee}$ in CF$_4$ gas at 76~Torr (0.1~atm) is less than 1~mm.
This is comparable to the strip pitch of 0.4~mm and one can deduce that the angular resolution and gamma-ray rejection both get worth below this point.
One solution is to operate the CF$_4$ gas at a lower pressure than 76~Torr to allow the nuclei and electrons run longer and improve the angular resolution and gamma-ray rejection below 50~keV$_{ee}$.

The remaining background sources are the ambient gamma-rays and internal radons as shown in Fig.~\ref{fig:spectrum}. 
We have already discussed the gamma-ray reduction above so we discuss the reduction of radon background here. 
The LA$\mu$-PIC, significantly reduced the surface alpha rays in NEWAGE2021, still contains some material which emanates the radon gas\,\cite{hashimoto_nim}. A new version of  the $\mu$-PIC series, LBG$\mu$-PIC currently being developed. The material used for the LBG$\mu$-PIC is  carefully selected so that the total radon emanation is less than 1/10 of the LA$\mu$-PIC.



With the improvements described above, we aim to explore the region claimed by DAMA/LIBRA\,\cite{DAMA} and to improve the sensitivity to reach limits by other direct search experiments.

\section{Conclusion}
A direction-sensitive direct dark matter search was carried out at Kamioka Observatory with a total
live time of 318.0 days corresponding to an exposure of 3.18 kg$\cdot$days.
A new gamma-ray rejection cut, which improved the gamma-ray rejection power to 8.8~$\times$~10$^{-7}$ while maintaining the detection-selection efficiency of the nuclear recoil at about 20\% was introduced.
This enabled us to use the high gas gain data, which was not used in the previous study due to the  deterioration of the gamma-ray rejection power. The exposure was increased by a factor of 2.4.
A 3D-vector reconstruction 
with a head-tail determination power of 52.4\% in the energy range of 50--100~keV was also used for this study.
As a result of the directional  WIMP-search analysis, 
an upper limit of the spin-dependent WIMP-proton cross section of 25.7~pb for a WIMP mass of 150~GeV/$c^2$ was derived. This limit marked the best direction-sensitive limit.


\section*{Acknowledgment}
This work was partially supported by KAKENHI Grant-in-Aids (19H05806, 19684005, 23684014, 26104005, 21K13943, 22H04574, and 21H04471).

\vspace{0.2cm}
\noindent
\let\doi\relax
\bibliographystyle{ptephy}
\bibliography{ref}

\end{document}